\documentclass{aastex631}

\usepackage{txfonts}
\usepackage{xspace}
\usepackage{mathrsfs}

\newcommand{\nnhp}{$\rm N_2H^+$\xspace}

\newcommand{\nndp}{$\rm N_2D^+$\xspace}
\newcommand{\hcop}{$\rm HCO^+$\xspace}
\newcommand{\dcop}{$\rm DCO^+$\xspace}
\newcommand{\co}{$\rm CO$\xspace}
\newcommand{\cdo}{$\rm C^{18}O$\xspace}
\newcommand{\hcdop}{$\rm HC^{18}O^+$\xspace}
\newcommand{\kms}{$\rm km \, s^{-1}$\xspace}

\newcommand{\tex}{$T_\mathrm{ex}$\xspace}
\newcommand{\vlsr}{$V_\mathrm{lsr}$\xspace}
\newcommand{\taup}{$\tau \rm  ^{peak}$\xspace}
\newcommand{\sigmav}{$\sigma_\mathrm{V}$\xspace}

\newcommand{\vin}{$V_\mathrm{infall}$\xspace}

\shorttitle{A large contracting envelope}
\shortauthors{Redaelli E.}
\graphicspath{{./}{figures/}}

\begin{document}

\title{A large ($\approx 1\, \rm pc$) contracting envelope around the prestellar core L1544\footnote{This work is based on observations carried out with the IRAM 30m telescope. IRAM is supported by INSU/CNRS (France), MPG (Germany) and IGN (Spain).}}

\correspondingauthor{Elena Redaelli}
\email{eredaelli@mpe.mpg.de}

\author[0000-0002-0528-8125]{Elena Redaelli}
\affiliation{Centre for Astrochemical Studies, Max-Planck-Institut f\"ur extraterrestrische Physik, Gie{\ss}enbachstra{\ss}e 1, 85749 Garching bei M\"unchen,
Germany}

\author{Ana Chac\'{o}n-Tanarro}
\affiliation{ Observatorio Astron\'omico Nacional (OAN/IGN), Alfonso XII, 3, 28014, Madrid, Spain}

\author{Paola Caselli}
\affiliation{Centre for Astrochemical Studies, Max-Planck-Institut f\"ur extraterrestrische Physik, Gie{\ss}enbachstra{\ss}e 1, 85749 Garching bei M\"unchen,
Germany}

\author{Mario Tafalla}
\affiliation{ Observatorio Astron\'omico Nacional (OAN/IGN), Alfonso XII, 3, 28014, Madrid, Spain}

\author{Jaime E. Pineda}
\affiliation{Centre for Astrochemical Studies, Max-Planck-Institut f\"ur extraterrestrische Physik, Gie{\ss}enbachstra{\ss}e 1, 85749 Garching bei M\"unchen,
Germany}

\author{Silvia Spezzano}
\affiliation{Centre for Astrochemical Studies, Max-Planck-Institut f\"ur extraterrestrische Physik, Gie{\ss}enbachstra{\ss}e 1, 85749 Garching bei M\"unchen,
Germany}

\author{Olli Sipil\"a}
\affiliation{Centre for Astrochemical Studies, Max-Planck-Institut f\"ur extraterrestrische Physik, Gie{\ss}enbachstra{\ss}e 1, 85749 Garching bei M\"unchen,
Germany}


\begin{abstract}
Prestellar cores, the birthplace of Sun-like stars, form from the fragmentation of the filamentary structure that composes molecular clouds, from which they must inherit at least partially the kinematics. Furthermore, when they are on the verge of gravitational collapse, they show signs of subsonic infall motions. How extended these motions are, which depends on how the collapse occurs, remains largely unknown. We want to investigate the kinematics of the envelope that surrounds the prototypical prestellar core L1544, studying the cloud-core connection. To our aims, we observed the \hcop (1-0) transition in a large map. \hcop is expected to be abundant in the envelope, making it an ideal probe of the large-scale kinematics in the source. We modelled the spectrum at the dust peak by means of a non local-thermodynamical-equilibrium radiative transfer. In order to reproduce the spectrum at the dust peak, a large ($\sim 1\, \rm pc$) envelope is needed, with low density (tens of $\rm cm^{-3}$ at most) and contraction motions, with an inward velocity of $\approx 0.05\,$\kms. We fitted the data cube using the Hill5 model, which implements a simple model {for the optical depth and excitation temperature profiles along the line-of-sight,} in order to obtain a map of the infall velocity. This shows that the infall motions are extended, with typical values in the range $0.1-0.2\,$\kms. Our results suggest that the contraction motions extend in the diffuse envelope surrounding the core, which is consistent with recent magnetic field measurements in the source, which showed that the envelope is magnetically supercritical.
\end{abstract}



\section{Introduction\label{intro}}
Low-mass stars form from the fragmentation of molecular clouds in dense ($n\gtrsim 10^5\, \rm cm^{-3}$) and cold ($T \lesssim 10\, \rm K$) cores, which are typically fraction of parsecs in size\footnote{See \cite{Pineda22} for an extensive review on the formation of structures in the interstellar medium.}. When the cores lack evidence of an embedded protostellar source, they are defined starless. A subset of starless cores is constituted by those that are gravitationally-bound, highly centrally-peaked, and on the verge of gravitational collapse. These, known as prestellar cores, represent the birthplace of Sun-like stars \citep{Andre00, Bergin07}. From the kinematic point of view, prestellar cores are quiescent structures, in contrast with the turbulent environment in which they are usually embedded \citep{Myers83,Goodman98,Caselli02c}, and they often show subsonic infall motions \citep{Tafalla98,Lee01}.
 \par
Several observational results showed that molecular clouds present a highly filamentary structure, as unveiled by
the \textit{Herschel} space telescope (see e.g. \citealt{Andre10}), in which prestellar cores are embedded \citep{Arzoumanian11, Palmeirim13, Kirk13}. These filaments often present a complex kinematics. They appear to be formed by fibers, i.e. coherent structures in space and velocity \citep{Hacar13,Tafalla15}.  From the theoretical point of view, it has been shown in simulations that fibers arise from the evolution of magnetised and
turbulent clouds under self-gravity \citep{ZamoraAviles17, Moeckel15}. Since cores are formed from these fibers, their kinematics could be at least partially inherited from the larger-scale structures. Observational studies focused on the core-cloud transition, especially from a kinematic point of view, could unveil important information on this connection, but so far they are scarce in literature. 
\par
On the other hand, the kinematics of dense cores and their surrounding envelope depend also on the properties of the gravitational contraction they are experiencing. For instance, the collapse of the isothermal sphere modelled by \cite{Shu77} occurs in an inside-out fashion. The fastest contraction motions are found towards the centre, whilst the outer envelope remains essentially static. On the contrary, in the Larson-Penston model \citep{Larson69, Penston69} the collapse happens outside-in, since small perturbations in the outskirts give rise to the initial contraction of the outer layers of the core, which then propagate to the central regions. More recently, it has been shown that observations of several cores are compatible with quasi-equilibrium contraction of a Bonnor-Ebert sphere (QE-BES; see e.g. \citealt{Keto15}, and references therein). Determining observationally how the infall velocity changes in the core/envelope structure allows to disentangle between the different collapse models.  \par
The observational investigation of contraction motions relies on the detection of molecular tracers with optically thick transitions, which present strong self-absorption at the rest velocity of the source. Inward motions result in the so-called blue asymmetry in the line profiles, where the blue (low velocity) peak is brighter than the red (high velocity) one  \citep{Leung77, Myers96, Evans99}. Typical candidates for this kind of observations are \nnhp, \hcop, or CS, among others. Most observational catalogs of contraction motions, however, consist of single-pointing observations at the cores' dust peaks \citep{Mardones97, Lee99,Lee01, Sohn07, Schnee13, Campbell16}. \cite{Keown16} performed one of the few spatially-resolved studies, as they investigated the infall motions of two prestellar cores (L694-2 and  L492). Those authors used grids of pointing observations of \dcop (1-0) and (3-2), \nnhp (1-0), and \hcop (3-2) transitions. They modelled the observed spectra by means of the Hill5 model \citep{deVries05}, which simulates {the line of sight properties of a} core contracting at constant velocity. The authors found typical infall velocity values of $0.05-0.15\,$\kms, with hints of decrease at radii $\approx 0.04\, \rm$pc from the dust peaks. The results however tell us about the properties of the denser cores more than of their surrounding envelopes, since \textit{i)} they report detections only within $0.1\, \rm pc$ from the dust peaks and  \textit{ii)} the targeted transitions either have high critical densities ($n_c \approx 10^6 \, \rm cm^{-3}$ for the (3-2) lines), or they are expected to trace only the very central parts of the cores due to chemical considerations (e.g., \nnhp). \par

L1544 is one of the best studied prestellar cores. Embedded in the Taurus molecular cloud at a distance of $d = 170 \, \rm pc$ (recently revised using GAIA data, \citealt{Galli19}), it has bright emission at millimetre wavelengths \citep{WardThompson99}, and it shows a rich and complex chemistry, as investigated for instance by \cite{Spezzano17}. In the central few thousands AU the core is cold, as seen with ammonia observations ($T_\mathrm{k}  < 7\, \rm K$, \citealt{Crapsi07}), and dense ($n > 10^6 \rm \, cm^{-3}$, \citealt{Tafalla02,Keto10,Caselli19}). 
The kinematics (and in turn the physical structure) of L1544 has been deeply investigated in the last decades, and yet it is far from being completely understood, due to its complexity. \cite{Myers96} and \cite{Tafalla98} firstly reported self-absorbed profiles in optically thick lines, such as $\rm HCO^+ $ (1-0) and $\rm CS$ (2-1), as well as double-peak profiles in optically thin lines, such as the $\rm C^{34}S$ (2-1)  and the $\rm H_2CO$ $ (2_{12}-1_{11})$ transition. The authors modelled the observations using a two-layer model: a denser, background layer responsible of the molecular emission, and a foreground one, with slightly different velocity, able to absorb sub-thermally excited lines. \cite{Caselli02a} further investigated this point using several ionised and neutral tracers. In that paper, the observations were modelled using two gas components, a denser one towards the core centre and a more diffuse one around it, which ---when interposed on the line-of-sight to the observer--- causes the absorption of the emission from the denser component. The authors however highlighted the importance of contraction (infall) motions to reproduce the observations, rather than distinct velocity components on the line-of-sight. \par   
\cite{Keto15} modelled the physical structure of the core in terms of density, temperature, and velocity field as a unidimensional quasi-equilibrium Bonnor-Ebert sphere (QE-BES), using $\rm C^{18} O$ and $\rm H_2O$ observations. The $\rm H_2O$ $(1_{10} - 1_{01})$ line, in particular, shows an inverse P-Cygni profile, consistent with contraction motions. The derived velocity field has a peak of $ -0.15 \rm \, km \, s^{-1}$ at $ R \approx 1000 \rm AU$. The model has been successfully used to reproduce several molecular line data using a non local-thermodynamic-equilibrium (non-LTE) radiative transfer approach (see e.g. \citealt{Caselli12,Bizzocchi13,Caselli17,Redaelli18}). In particular, in \cite{Redaelli19} we used the QE-BES model in combination with the abundance profiles resulting from a chemical code \citep{Sipila19} to fit the emission at the L1544 dust peak of several rotational lines of $\rm HC^{18}O^+$, $\rm DCO^+$, $\rm N_2H^+$, and $\rm N_2D^+$, as well as the HCO$^+$ (1-0) spectrum from \cite{Tafalla98}. The chemical model that provided the best agreement to the observations of the \nnhp isotopologues and $\rm DCO^+$ was characterised by an external visual extinction (which simulates the presence of the surrounding cloud) of $A_\mathrm{V} = 1$ or $2 \, \rm mag$. On the contrary, this value had to be increased to $A_\mathrm{V} = 5 \, \rm mag$ to reproduce the  HCO$^+$ observations. This stronger shielding increases the molecule abundance in the core's outskirts, and it allows to obtain at least partially the blue asymmetry. \par
It is of great interest to investigate spatially the contraction motions, in order to understand how extended they are, and if they show spatial variations, in the envelope of such a well-known source. In order to do so, we discuss a large ($7'\times6'$, corresponding to $0.35\rm pc \times 0.30 pc$ at L1544 distance) map of \hcop (1-0) line towards L1544. This line has a critical density of $n_c =7\times  10^4 \, \rm cm^{-3}$, but its effective critical density can be as low as  $10^3\, \rm cm^{-3}$ at $10 \, \rm K$, and it decreases at higher temperatures \citep{Shirley15}. in fact, its high optical depth contributes to lowering the equivalent critical density of \hcop, due to photon trapping. Moreover, \hcop, which is formed mainly from CO, is expected to suffer from depletion at the high densities at the core's centre, whilst it is still abundant in the more diffuse material surrounding it. Indeed, \hcop is present also in diffuse clouds (see for instance \citealt{Liszt94, Lucas96}). All these considerations make this line an ideal probe of the properties of the large-scale envelope of L1544.  \par
The paper is organised as follows. The observational setup is presented in Sect. \ref{obs}, and the resulting data are shown in Sect. \ref{res}. Section \ref{disc} reports the analysis and discussion of the data, where first we model the observations at the dust peak using a non-local-thermodynamical-equilibrium approach, and then we fit the whole data cube using the Hill5 model from \cite{deVries05}. Finally, Sect. \ref{conclusion} contains a summary and conclusions of this work. 

\section{Observations\label{obs}}
We observed the \hcop (1-0) transition at $89.18852 \, \rm GHz$\footnote{From the CDMS catalog, \url{www.cdms.astro.uni-koeln.de}.} with the Institut de Radioastronomie Millim\'etrique (IRAM) 30m telescope, located at Pico Veleta (Spain) in August 2021. We used the E090 band of the Eight MIxer Receiver (EMIR) as a frontend, combined with the VErsatile SPectrometer Assembly (VESPA) backend, with a spectral resolution of $20\, \rm kHz$ (corresponding to a velocity resolution of $\approx 0.065 \,$\kms). The chosen observation mode was on-the-flight (OTF) mapping, covering an area of $7'\times6'$, Nyquist-sampled, with position switching. The weather presented average condition for summer months, with a typical precipitable water vapour of $PWV = 7-8 \, \rm mm$, corresponding to an atmosphere opacity of $\tau = 0.4-0.5$ at $225\, \rm GHz$. Uranus and the bright quasar 0316+413 were used to perform focus observations. Pointing was also checked on the quasars 0439+360 and 0528+134 every $1.5-2\, \rm h$, and found to be generally accurate within $4''$. \par
The data were reduced using the GILDAS package\footnote{Available at \url{http://www.iram.fr/IRAMFR/GILDAS/}.},
and they were calibrated into main beam temperature ($T_\mathrm{MB}$) from the antenna temperature scale using $T_\mathrm{MB} = T_\mathrm{A}^* \times F_\mathrm{eff} / B_\mathrm{eff}$, where the beam efficiency is $B_\mathrm{eff} = 0.81$ and the forward efficiency is $F_\mathrm{eff} =0.95$. The final angular resolution is $29''$, corresponding to $\approx 5000 \, \rm AU$ at the source distance. The mean sensitivity of the final data cube is $rms = 100 \, \rm mK$ in the $0.07\,$\kms channel, computed on the line-free channels. \par
We will also discuss the emission of the \cdo (2-1) line at $\rm 219.56035 \, GHz$ towards L1544. These data will be fully discussed in an upcoming paper (Chac\'on-Tanarro et al., in prep), but we provide here some information. The line was observed with the IRAM 30m telescope during 2014 with the E230 band of EMIR. OTF mapping with position switching was used, covering a large footprint ($\approx 150 \, \rm arcmin^2$). Here we focus only on the part of the FoV overlapping the one of the \hcop data. The VESPA backend was used, with a spectral resolution of $20\, \rm kHz$, corresponding to a velocity resolution of $\approx 0.026 \,$\kms. The beam efficiency and forward efficiency at $217 \, $GHz are  $B_\mathrm{eff} = 0.61$ and  $F_\mathrm{eff} =0.92$, respectively, and the beam size is $11.8''$.
   \begin{figure*}[!t]
    \centering
   \includegraphics[ width=.8\textwidth]{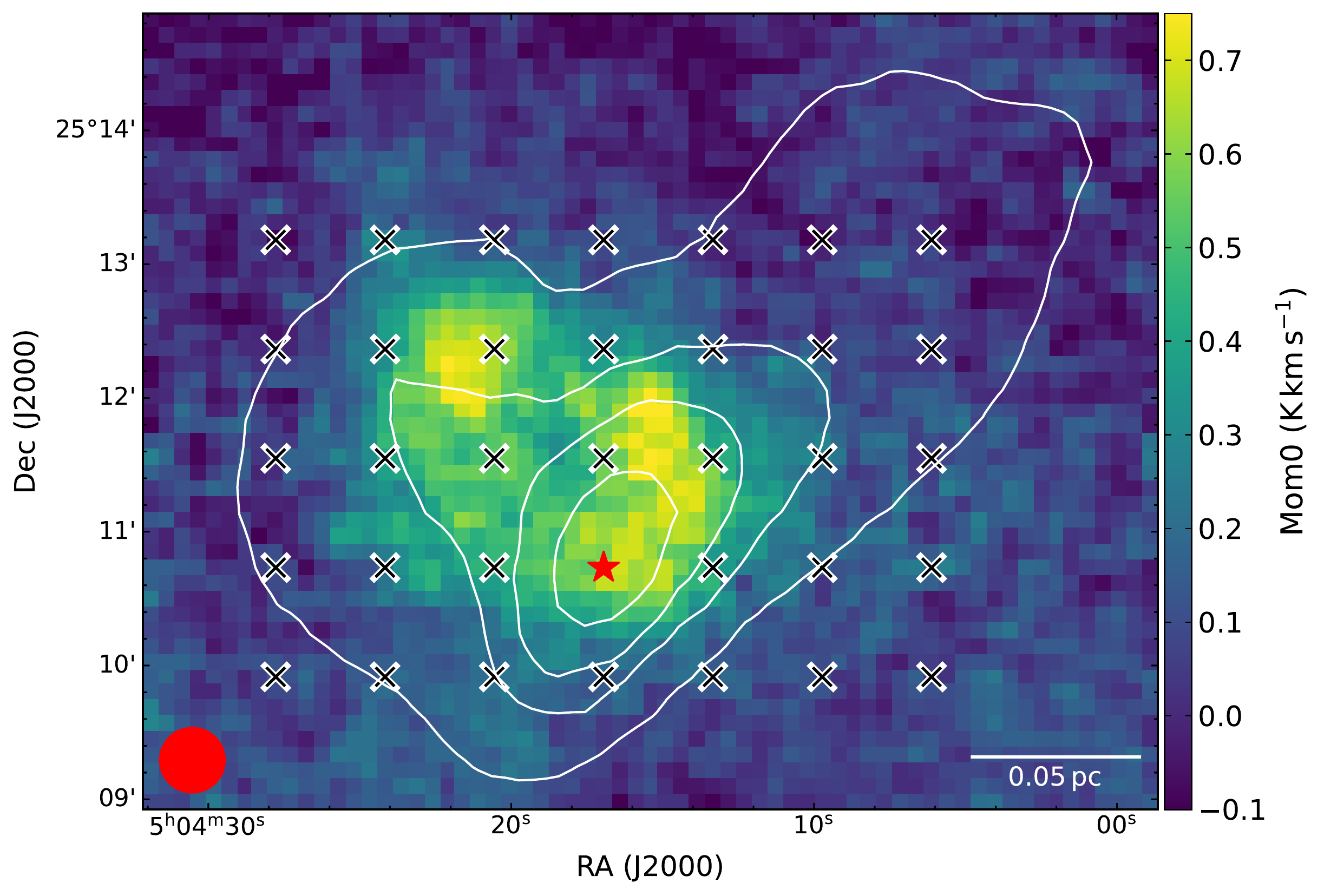}
      \caption{The integrated intensity of \hcop (1-0) is shown in colorscale (mean $rms=40\, \rm mK\,$\kms) . The white contours show the $N\rm (H_2)$ distribution at levels $[5,9,13,20]\times 10^{21} \, \rm cm^{-2}$, taken from \cite{Spezzano17}. The red star shows the millimetre dust peak position ($\rm RA(J2000)= 05^h 04^m 17^s.21$, $\rm Dec(J2000) = 25^d 10^m 42^s.8$, \citealt{WardThompson99}). The crosses represent the positions where the spectra of Fig. \ref{SpectraGrid} have been extracted. The IRAM beam size (29'') and scalebar are shown in the bottom left and right corners, respectively.  \label{mom0}}
   \end{figure*}

\section{Results\label{res}}
Figure \ref{mom0} shows the integrated intensity map of the \hcop (1-0) transition, computed in the velocity range $6-8\,$\kms, overlaid with contours representing the H$_2$ column density obtained from \textit{Herschel} data. In dust thermal emission, the core appears v-shaped, with two filamentary structures at column densities of $2-4\times 10^{21} \, \rm cm^{-2}$ elongated towards north-west and north-east. In the south-west direction, the core presents a sharp decrease of the gas column density, in correspondence with the edge of the Taurus cloud. The integrated intensity of \hcop traces the gas in the opposite direction of this steep edge with respect to the dust peak. Its morphology shows two separate peaks, one towards the north-west direction from the dust peak (but connected to it), and the second in the {gas} structure {that elongates} towards north-east. We highlight that the {high opacity} of the \hcop line affects also its integrated intensity, which does not reflect the distribution of the molecular column density. This explains why the morphology of the \hcop (1-0) lines shown in Fig.~\ref{mom0} is significantly different from that of the integrated flux of the optically thin isotopologue \hcdop (see Fig. 3 in \citealt{Redaelli19}).
 \begin{figure*}[!h]
    \centering
   \includegraphics[angle = -90 , width=.8\textwidth]{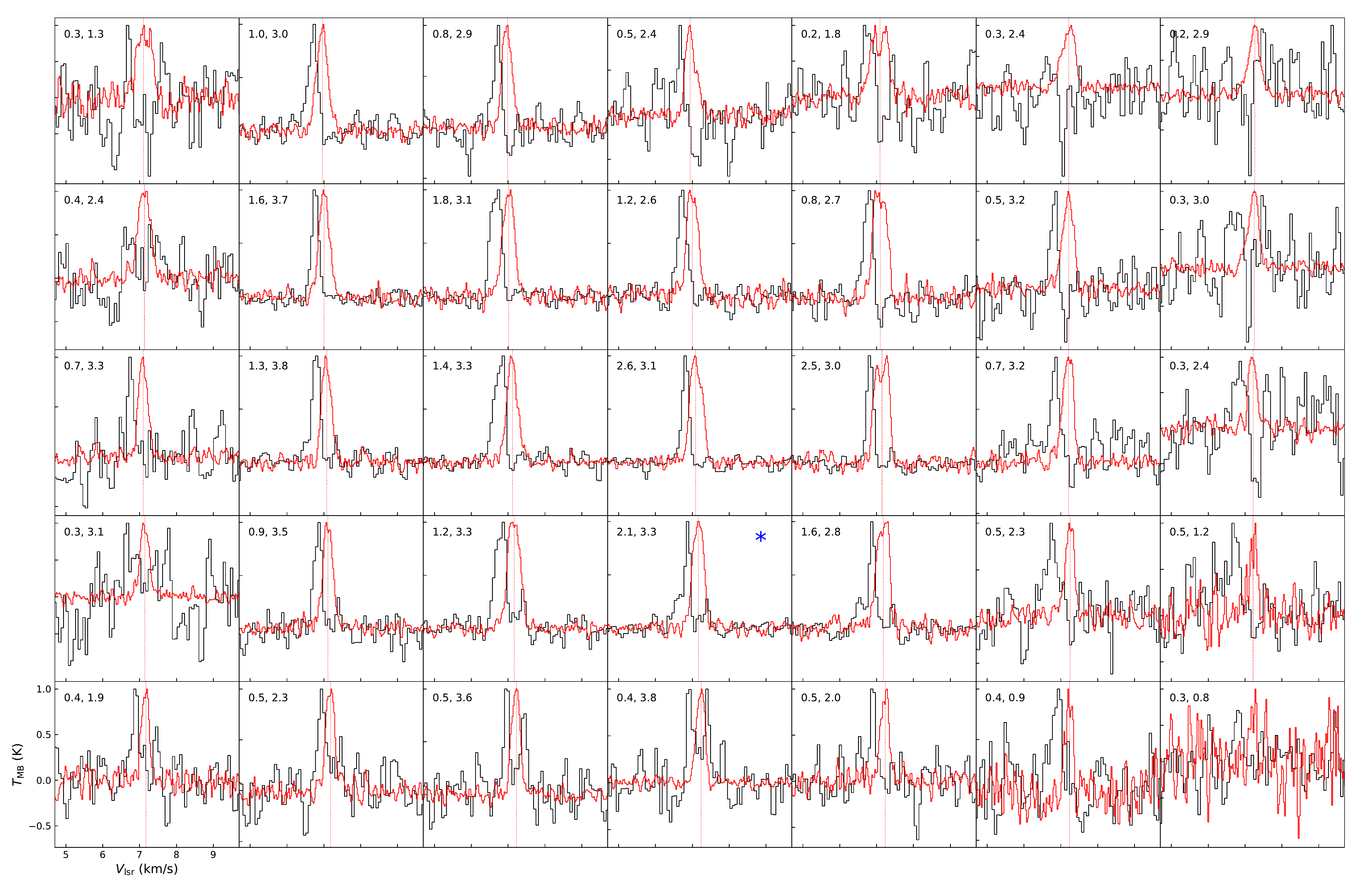}
      \caption{The black histograms are the \hcop spectra at the positions shown with black crosses in Fig. \ref{mom0}, which are separated by $49''$ in both right ascension and declination. The panel with the blue asterisk represents the spectrum at {the} millimetre dust peak. The red histograms show the corresponding \cdo (2-1) spectra. {The spectra of both molecules are normalised to their peak values, which are indicated in the top-left corner of each panel (in K, first for \hcop, and then for \cdo).} The vertical {dotted line (in red)} shows the centroid velocity obtained fitting a single Gaussian profile to the \cdo data. {The same figure in absolute flux units is reported in Appendix \ref{UnscaledSpectra}.}
         \label{SpectraGrid}}
   \end{figure*}
\section{Analysis and discussion\label{disc}}

\subsection{An extended, contracting envelope\label{spectra_descript}}
Figure \ref{SpectraGrid} shows with black histograms a grid of \hcop (1-0) spectra taken at intervals of $49''$ (approximately 1.5 times the beam size), covering the whole map (see Fig. \ref{mom0} for the positions where the spectra are extracted). {The spectra are shown in scaled units, to better appreciate their profiles.} The data present double peak features, with strong blue asymmetries, especially around the dust peak. In order to exclude that these profiles are to be attributed to multiple velocity components on the line of sight, we have compared the \hcop spectra with \cdo (2-1) observations (Chac\'on-Tanarro et al, in prep). In order to allow for a proper comparison, the \cdo datacube has been smoothed to the resolution of the \hcop one, and it has been re-gridded to the same coordinate grid. {The targeted \cdo transition is only moderately optically thick ($\tau \lesssim 1.3$)\footnote{ {We evaluate the maximum optical depth using $\tau = -\ln \left [1- \frac{T_\mathrm{MB}}{J_\nu(T_\mathrm{ex}) - J_\nu(T_\mathrm{bg})} \right]$, where $J_\nu$ is the equivalent Rayleigh-Jeans temperature, $T_\mathrm{MB}=4 \ $K is the peak intensity of the \cdo transition, $T_\mathrm{bg}=2.73 \ $K is the background temperature, and  $T_\mathrm{ex}=10 \ $K is the line excitation temperature. We assume this value since it roughly corresponds to the gas temperature at the position in the core where CO depletion starts (see Fig.~\ref{PhysModel} and \ref{MOLLIEco}), under the assumption that the line is thermalised.}} even in the densest part of the source. As a consequence, its line profile is not expected to be affected by self-absorption (and in fact this feature is not seen in the spectral profiles, see Fig. \ref{SpectraGrid}). The centroid of this line is hence expected to trace the source velocity.} \par
The \cdo (2-1) spectra are shown in Fig. \ref{SpectraGrid} with red histograms. In all positions where both lines are detected, the peak of the \cdo line is red-shifted with respect to the \hcop (1-0) peak. At those positions where \hcop shows an asymmetric, double-peaked profile, the \cdo line sits right in the middle between the two peaks. We fit a single Gaussian line to the \cdo data cube pixel-by-pixel, using the \textsc{python pyspeckit} package \citep{Ginsburg11}. This yields the map of the line {local-standard-of-rest-velocity} $V_\mathrm{lsr}$. In each panel of Fig. \ref{SpectraGrid} the derived \vlsr value is shown with a dashed, vertical line. This is always found to correspond with the dip between the two peaks of the \hcop line. \par
This comparison leads to the conclusion that the spectral profiles of the \hcop (1-0) line are not due to multiple components on the line of sight, but instead they are caused by {a combination of internal motions within the source and self-absorption due to line opacity effects.} Interestingly, the velocity shift between the \hcop and the \cdo line is visible also towards those positions where the former seems to present a single-peaked profile (e.g. towards the north-east of the map coverage), hinting that these spectra are affected by such strong self-absorption, that the red peak is completely absorbed. Our data hence support a scenario where the whole envelope encompassed by the map coverage is experiencing contraction, due to the gravitational pull towards the centre. \par
   
    \begin{figure*}[!t]
    \centering
   \includegraphics[width=.7\textwidth]{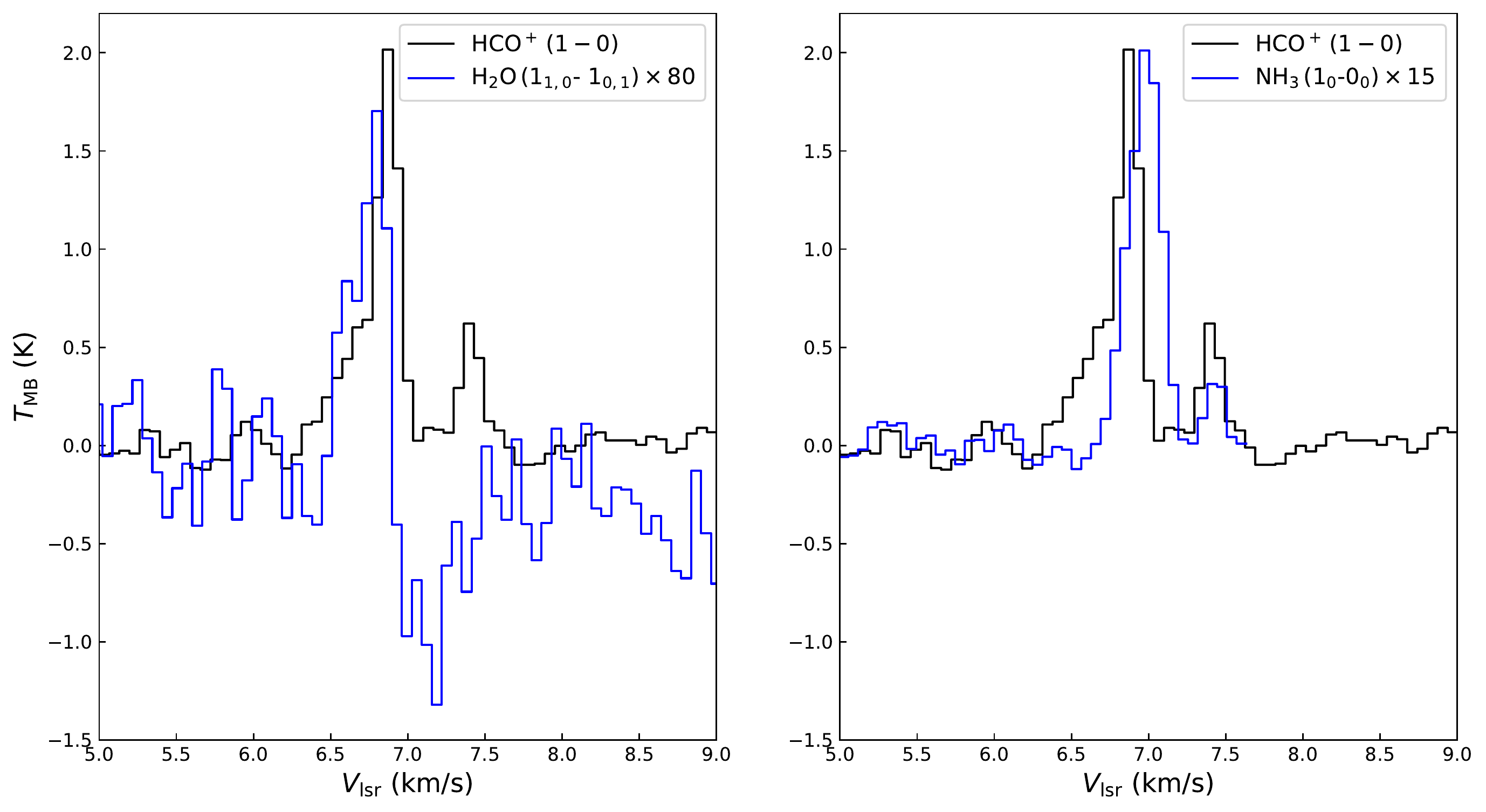}
      \caption{ \textit{Left panel:} comparison between the \hcop (1-0) spectrum (black histogram) and the ortho-$\rm H_2O (1_{1,0}- 1_{0,1})$ (blue curve) spectrum observed by \textit{Herschel}/HIFI \citep{Caselli12}, taken at the millimetre dust peak of L1544. The water line has been scaled up by a factor of 80 to allow an easier comparison. \textit{Right panel:} same as in the left panel, but this time the blue histogram shows the isolated hyperfine component $F'_\mathrm{N} \rightarrow F_\mathrm{N}= 0 \rightarrow 1$  of the ortho-$\rm NH_3 (1_0-0_0)$ line (from \citealt{Caselli17}), and it has been scaled by a factor of 15. All the spectra have been smoothed to the same resolution of $40''$. Note the spectral resolution of the HIFI instrument at the frequencies of the $\rm H_2O$ and $\rm NH_3$ lines is $\approx 0.065 \,$\kms, hence comparable to that of the \hcop data. Both panels are shown with the same y-axis range. \label{SpectraComp}}
   \end{figure*}
   
  Figure \ref{SpectraComp} shows a comparison of the \hcop (1-0) spectrum at the L1544 dust peak with the ortho-$\rm H_2O (1_{1,0}- 1_{0,1})$ line (left panel, from \citealt{Caselli12}) and the ortho-$\rm NH_3 (1_0-0_0)$ line (right panel, from \citealt{Caselli17}). Both the water and the ammonia data were observed with the \textit{Herschel}/HIFI instrument. The \textit{Herschel} beam is $40''$, and we smoothed the \hcop data to this resolution before the comparison. \par
 As discussed in \cite{Caselli12}, the water line presents an inverse P-Cygni profile, a characteristic sign of gravitational contraction. In fact, it arises when the material in the back layers of the source is infalling towards the centre, hence emitting at blue-shifted velocities, whilst the bulk of the water gas in the foreground layers totally absorb the water emission from the dense regions of the core. The absorption is so strong that it is also seen against the faint continuum emission. The peak of the \hcop (1-0) line appears shifted towards higher velocities (even though only by one channel) with respect to the water line, likely due to the more severe self-absorption that affects $\rm H_2O$. \par
\cite{Caselli17} briefly discussed the striking similarities between the \hcop and the ortho-$\rm NH_3 (1_0-0_0)$ lines, but they used a \hcop (1-0) spectrum observed with the FCRAO-14m telescope \citep{Tafalla98}, which had a worse angular resolution (by approximately a factor of 2) and a worse sensitivity. With our new data, we confirm that the two transitions present similar blue-asymmetric line profiles, with a central dip that reaches the zero flux level. However, the \hcop line appears even more blue-shifted, presenting a blue excess that is instead similar to the one exhibited by the water line. \par
The origin of this blue excess remains difficult to explain. If it is related to a higher velocity component, it should arise from regions closer to the core centre. However the critical density of \hcop (1-0) is more than two orders of magnitude lower than that of ortho-$\rm H_2O (1_{1,0}- 1_{0,1})$ and ortho-$\rm NH_3 (1_0-0_0)$ ($n_\mathrm{c} \approx 10^7 \, \rm cm^{-3}$ for both, \citealt{Caselli12,Caselli17}). As a consequence, this feature should be present also in the  $\rm NH_3 (1_0-0_0)$, unless it is below the sensitivity level of the \textit{Herschel} spectrum. Furthermore, our current physical model does not present such high velocity values. This could be however related to the limitations of our one-dimensional model (see below). Ongoing work on a full 3D physical model, which takes into account asymmetric infall profiles, shows hints of higher infall velocities \citep{Caselli22}.

\subsection{Non-LTE modelling at the dust peak}
\begin{figure}[!h]
    \centering
   \includegraphics[width=.55\textwidth]{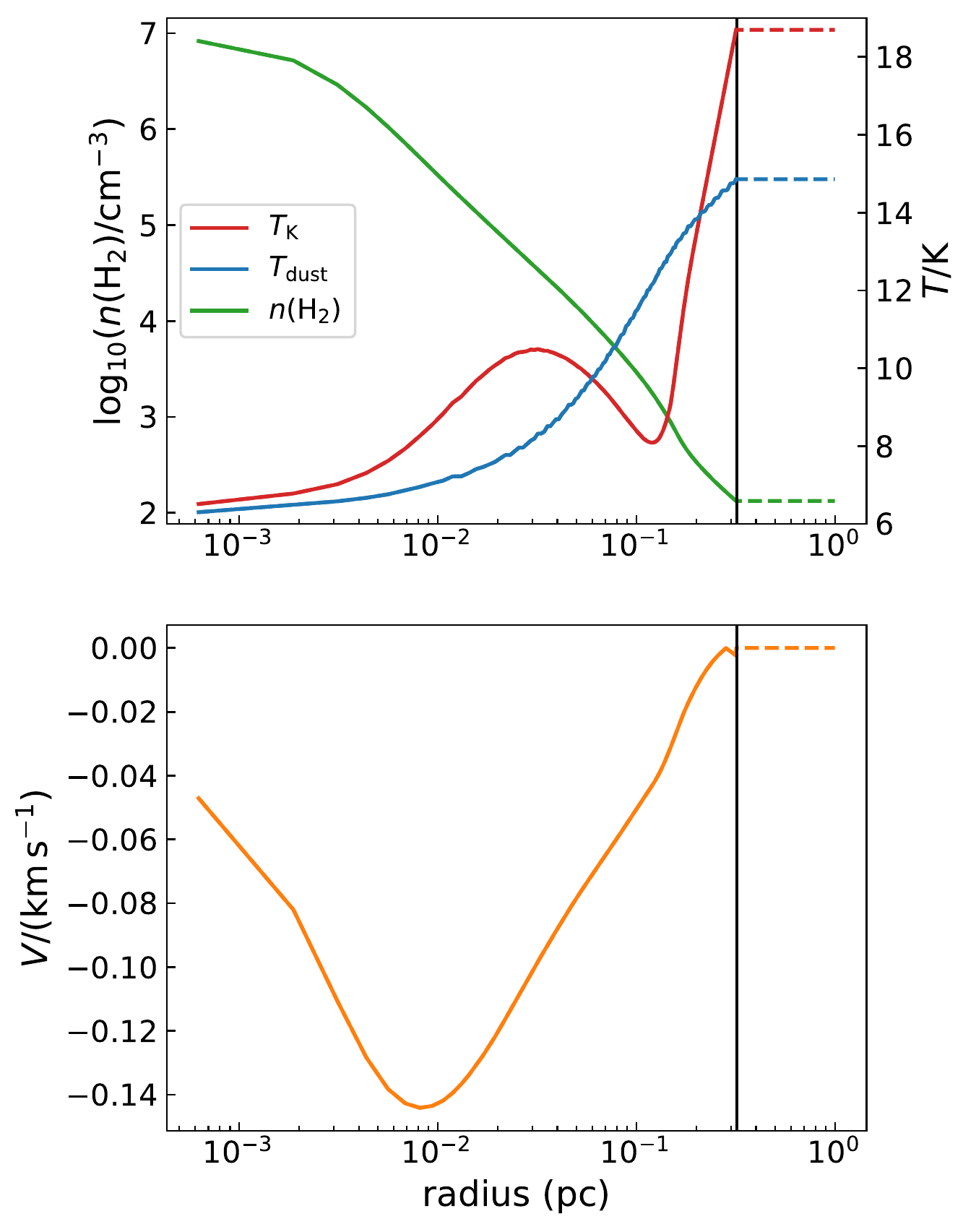}
      \caption{{\textit{Top panel:}} The solid curves show the original physical model computed by \cite{Keto15}: {H$_2$} volume density profile ({green} curve), gas temperature profile (red curve), and dust temperature profile (blue curve). The dashed extensions of the curve represent the properties of the isothermal, uniform envelope described in \ref{MOLLIEhcop} that extends from $ r = 0.32 \, \rm pc$ (shown with the vertical black line) to $1.0 \, \rm pc$. {\textit{Bottom panel:} same as the top, but for the velocity profile, shown in orange.} \label{PhysModel}}
   \end{figure}

In order to further investigate the properties of the contracting envelope, we first focus on modelling the observed \hcop spectrum at the dust peak with a non-LTE analysis, which has proven to be a powerful method to investigate the physical and chemical structure of L1544, already applied to several molecular species \citep{Caselli12, Caselli17, Bizzocchi13, Redaelli18, Redaelli19, Redaelli21}. This approach makes use of non-LTE radiative transfer code MOLLIE \citep{Keto90, Keto04}, coupled with the physical model of L1544 developed by \cite{Keto15}. This, based on a contacting Bonnor-Ebert sphere, consists of the gas density profile $n$, the dust ($T_\mathrm{dust}$) and gas ($T_\mathrm{K}$) temperature profiles, and the velocity profile $V$. The physical model we adopt here extends to $0.32\, \rm$pc, and it is shown in Fig. \ref{PhysModel} (solid lines). \par
Since it is one-dimensional and it has been constrained with spectroscopic observations of \cdo, \nnhp, and $\rm H_2O $ at the dust peak of L1544, the model is suitable to perform the radiative transfer analysis only at the core's centre. However, it can still provide important constraints on the radial physical structure and on the radial distribution of different species along the line-of-sight going through the dust peak. With this in mind, we use this approach to investigate the properties of the contracting gas that constitute the envelope of L1544. \par
   
The non-LTE radiative transfer requires the abundance profile $X_\mathrm{mol} = n_\mathrm{mol}/ n \rm (H_2)$ of the analysed species to compute synthetic spectra. To this aim, we use the gas-grain chemical model discussed in \cite{Sipila15a, Sipila15b, Sipila19}{, which implements photodissociation reactions from the KIDA database (\citealt{Wakelam12};  \url{http://kida.astrophy.u-bordeaux.fr}), and secondary photodissociation or ionisation reactions, with rates taken from \cite{Heays17}.} The chemical evolution in L1544 is simulated by dividing the physical model of \cite{Keto15} into concentric shells. The combination of the results in each shell at a given time step gives the radius-dependent abundance profiles of the analysed species. \cite{Redaelli21} used this approach applied to several transitions of \nnhp, \nndp, \dcop, and the optically thin \hcdop to investigate the cosmic-ray ionisation rate ($\zeta$) in L1544, in particular in terms of cosmic ray attenuation \citep{Padovani18}. The authors found that the model that provides the best agreement with the observational data is the ``low-model" of \cite{Padovani18}, which is consistent with the most recent Voyager data. In this model, the cosmic-ray ionisation rate profile decreases from $4\times 10^{-17} \, \rm s^{-1}$ at the edge of the core to $2\times 10^{-17} \, \rm s^{-1}$ in the central part, with an average value of $\langle \zeta \rangle = 3\times 10^{-17} \, \rm s^{-1}$. Throughout this paper, we will adopt the chemical model obtained adopting this profile for the cosmic-ray ionisation rate. We will discuss the solution at an evolutionary stage of $1\, \rm Myr$, which in \cite{Redaelli21} provided the best agreement to the observations\footnote{In \cite{Redaelli21}, the evolutionary timescale $t = 1\, \rm Myr$ provides the best fit solution for all investigated species, but \nndp. However, this tracer is a late-type molecule (because it is formed from molecular nitrogen), and furthermore it is deuterated, and it is very different from \hcop from a chemical point of view.}.

\subsubsection{\hcop modelling\label{MOLLIEhcop}}
As mentioned in Sect. \ref{intro}, \cite{Redaelli19} used the MOLLIE radiative transfer code, combined with the chemical model from \cite{Sipila19}, to model the \hcop (1-0) spectrum of \cite{Tafalla98}\footnote{The main difference from the modelling performed in \cite{Redaelli21} is the cosmic-ray ionisation rate, which in \cite{Redaelli19} is assumed to be constant and equal $\zeta = 1.3 \times 10^{-17} \, \rm s^{-1}$.  \cite{Redaelli21} did not analyse any \hcop data, but only the optically thin \hcdop isotopologue.}. In that paper we found that in order to reproduce at least partially the strongly asymmetric double peak profile of this line, the chemical model needed to use an external visual extinction of $A_\mathrm{V} = 5\, \rm mag$, instead of $A_\mathrm{V} = 1\, \rm mag$ which produced the best agreement for the other targeted species. This parameter simulates the external cloud embedding the core, and it regulates the attenuation of the UV flux impinging on the core. A higher
$A_\mathrm{V} $ value reduces the molecular photodissociation, and hence it provides a high abundance $X_\mathrm{mol} \approx 10^{-8}-10^{-7} $ also in the external layers of the core model, which are responsible for the self-absorption of the line emission. \par
\begin{figure}[h]
    \centering
   \includegraphics[width=.48\textwidth]{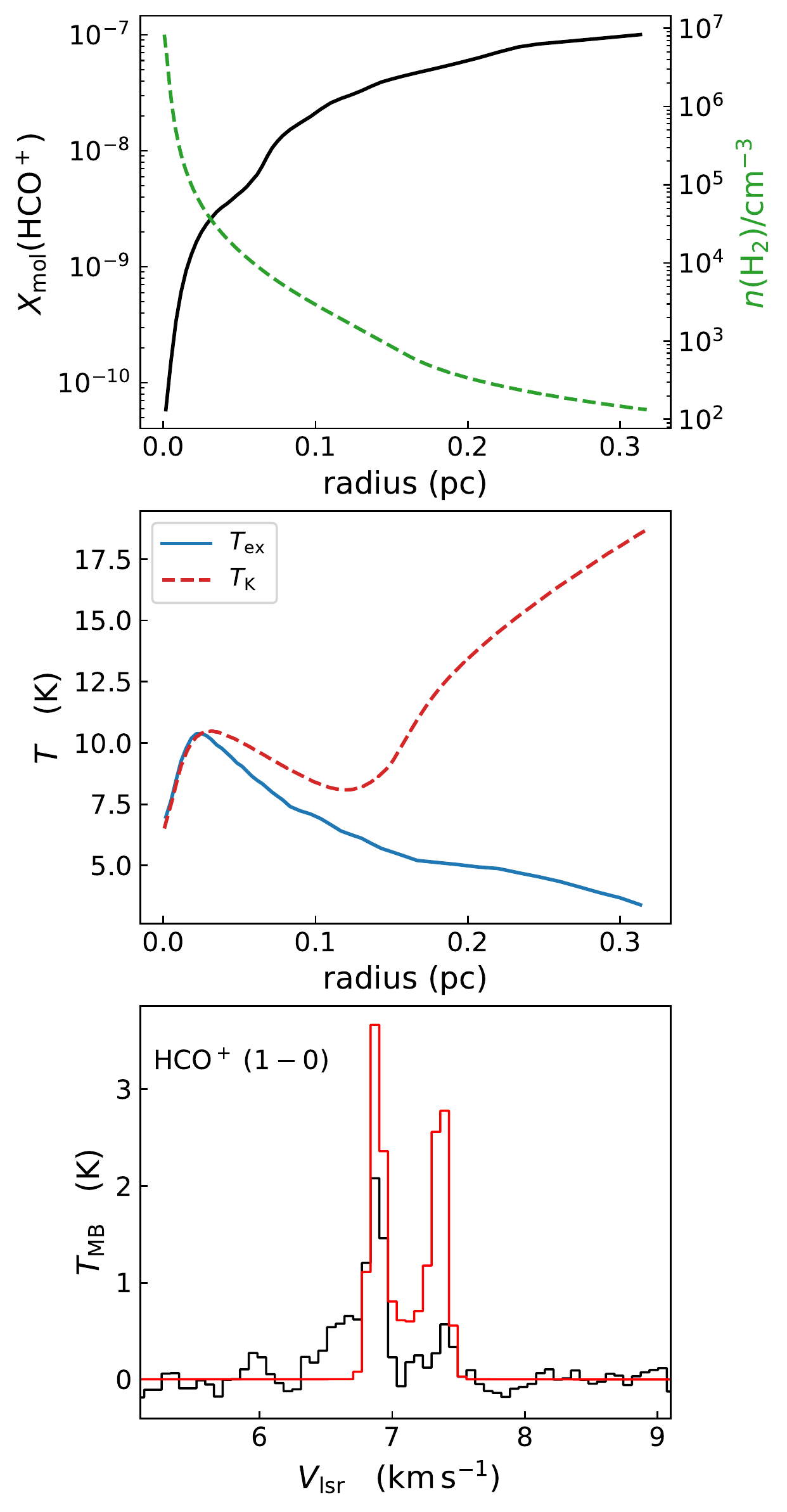}
      \caption{Input quantities and radiative transfer results for the initial model with no added envelope. {\textit{Top panel:}  The solid black curve shows the \hcop abundance profile predicted by the chemical model at $t=10^6 \, \rm yr$ and with an external visual extinction $A_\mathrm{V} = 5\, \rm mag$, compared with the H$_2$ volume density profile taken from the physical model (dashed green curve; both quantities are shown in logarithmic scale). \textit{Middle panel:} Comparison between the gas kinetic temperature of the model (dashed red curve) and the the excitation temperature profile of the (1-0) transition computed by MOLLIE (solid blue line).
      \textit{Bottom panel:} the observed \hcop (1-0) spectra towards the L1544 dust peak (black histogram) compared with the synthetic spectrum computed with MOLLIE using the abundance profile shown in the top panel (red histogram). }  \label{MOLLIEnoenv}}
   \end{figure}

The best-fit model reported by \cite{Redaelli19} failed however in reproducing two key features of the spectral profile of the \hcop line: \textit{i)} the red peak is suppressed with respect to the blue one only by a factor of $\approx 1.6$, whilst in the observed spectra the blue peak is 4 times brighter than the red one; \textit{ii)} the central dip flux does not reach the zero level. We tested the most recent chemical model (with the updated cosmic-ray ionisation rate profile) in the same conditions, and compared the simulated spectrum with the one extracted at the dust peak from our new IRAM data. Figure \ref{MOLLIEnoenv} shows the derived molecular abundance profile (top panel). \hcop is affected by strong depletion at high densities, and its abundance drops by three orders of magnitude from the core's external layers towards the centre. \par
The bottom panel of Fig. \ref{MOLLIEnoenv} shows the comparison of the observed and synthetic spectra. Similarly to \cite{Redaelli19}, the model fails in reproducing the zero level dip and the strong blue/red asymmetry. Furthermore, the simulated spectrum in general overestimates the line flux. The model does not produce enough self-absorption to suppress the red peak and to bring the flux between the peaks to zero. In order to have such a strong absorption, a layer of low-density gas, still rich in \hcop, is needed, where the molecule is not collisionally excited (i.e. it does not emit), and the excitation temperature ($T_\mathrm{ex}$) of the transition is close to the background temperature ($2.73\, \rm K$). The \tex radial profile of the model, which is provided as an output of MOLLIE, is also shown in the top panel of Fig. \ref{MOLLIEnoenv}. The curve peaks at $T_\mathrm{ex} \approx 10 \, \rm K$ at $0.025\, \rm pc$ (approximately where the gas temperature has its first peak, see Fig. \ref{PhysModel}), and it then decreases, decreasing below $4 \, \rm K$ beyond a radius of $0.29 \, \rm pc$. The minimum value it reaches is $3.3 \, \rm K$. 
\par

In order to improve the line modelling, we test the simulation of an extended envelope around the core. To do so, we use a simple approach, by extending the physical model of \cite{Keto15} from $0.32 \, \rm pc$ to $1.0 \, \rm pc$. In our model, this external $0.68 \, \rm pc$ envelope is hence uniform and isothermal, and its physical properties are equal to those at the edge of the initial, core-only model: the volume density is $n \rm _{env}   = 133 \, cm^{-3} $, the gas temperature is $T \rm _{K} ^{env}   = 18.7\, K  $, and the dust temperature is $T \rm _{dust} ^{env}   = 14.9\, K  $. The \hcop abundance is constant, and equal to $X_\mathrm{mol} \rm (HCO^+) = 10^{-7}$. In this first test, the envelope is static ($V_\mathrm{env} = 0\,$\kms ). Figure \ref{PhysModel} shows the properties of the core+envelope model. We highlight how this external layer accounts for a visual extinction of less than $0.3 \, \rm mag$. Its effect of attenuation of the external radiation field is limited, and we assume here that the temperature profile of the original model in the inner $0.32\, \rm pc$ is not affected.
\par
    \begin{figure*}[!t]
    \centering
   \includegraphics[width=\textwidth]{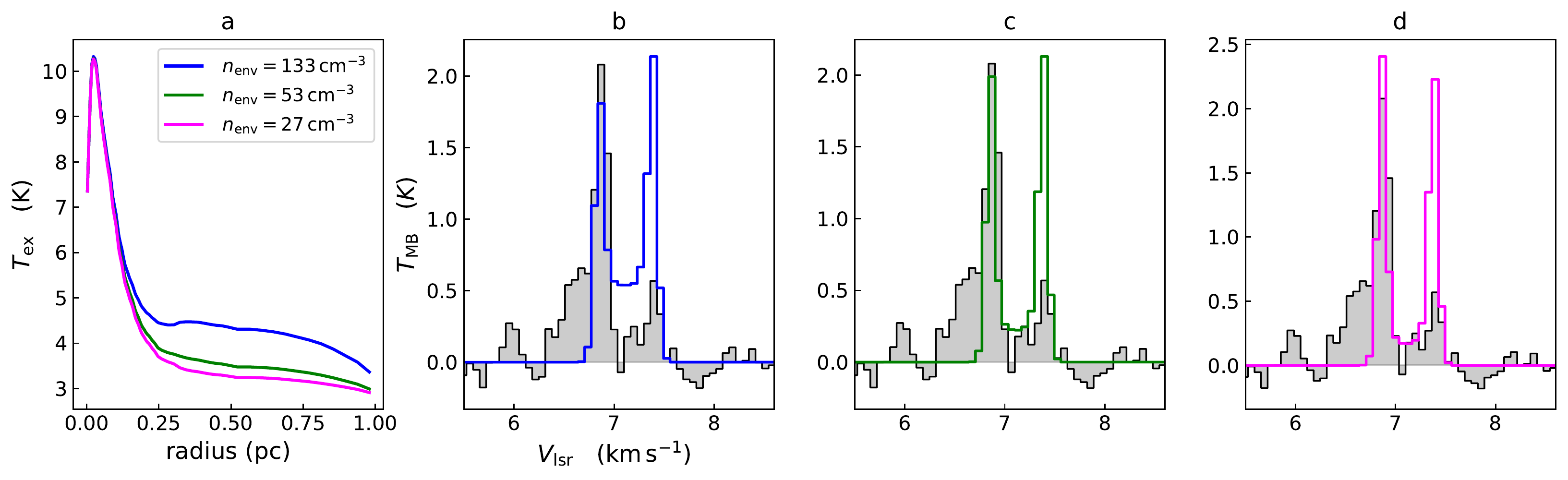}
      \caption{ \textit{Panel a):} Excitation temperature profiles of the \hcop (1-0) transition obtained with the model with the extended, static envelope, at three different volume densities: $n \rm _{env}   = 133 \, cm^{-3} $ (blue curve), $n \rm _{env}   = 53 \, cm^{-3} $ (green curve), and $n \rm _{env}   = 27  \, cm^{-3} $ (red curve). \textit{Panels b to d):}. Synthetic spectra obtained with the radiative transfer for the corresponding models presented in the panel a). The observed spectrum is shown with the black histogram and grey shadow in each panel.  \label{MOLLIEstaticenv}}
   \end{figure*}
   
Figure \ref{MOLLIEstaticenv} presents the synthetic spectra obtained from MOLLIE (second panel from the left). This model does not improve the agreement with the observations. In the extended envelope ($r>0.32\, \rm pc$), that has a density higher than $100 \, \rm cm^{-3}$, the line presents an average excitation temperature of $T_\mathrm{ex} = 4.2\, \rm K$ (left panel of Fig. \ref{MOLLIEstaticenv}). The transition is hence excited, and as a result, the absorption dip does not reach the zero level, and the blue wing is slightly less bright than the red one. In the assumption that the line is collisionally excited, the main parameter that regulates \tex is the gas density. We hence performed two more tests, decreasing  $n \rm _{env}$ respectively by a factor of $2.5$ and 5.  Figure \ref{MOLLIEstaticenv} shows the MOLLIE results also from these two models. As the envelope density decreases, the \tex value in the external layers also decreases. As a consequence, the flux dip between the two peaks approaches the zero level, in agreement with the observations. Furthermore, also the ratio between the intensity of the blue peak and the red one decreases, and in the model with $n \rm _{env}   = 27  \, cm^{-3} $ it becomes larger than one, as observed. \par

\begin{figure*}[t]
    \centering
   \includegraphics[width=1.\textwidth]{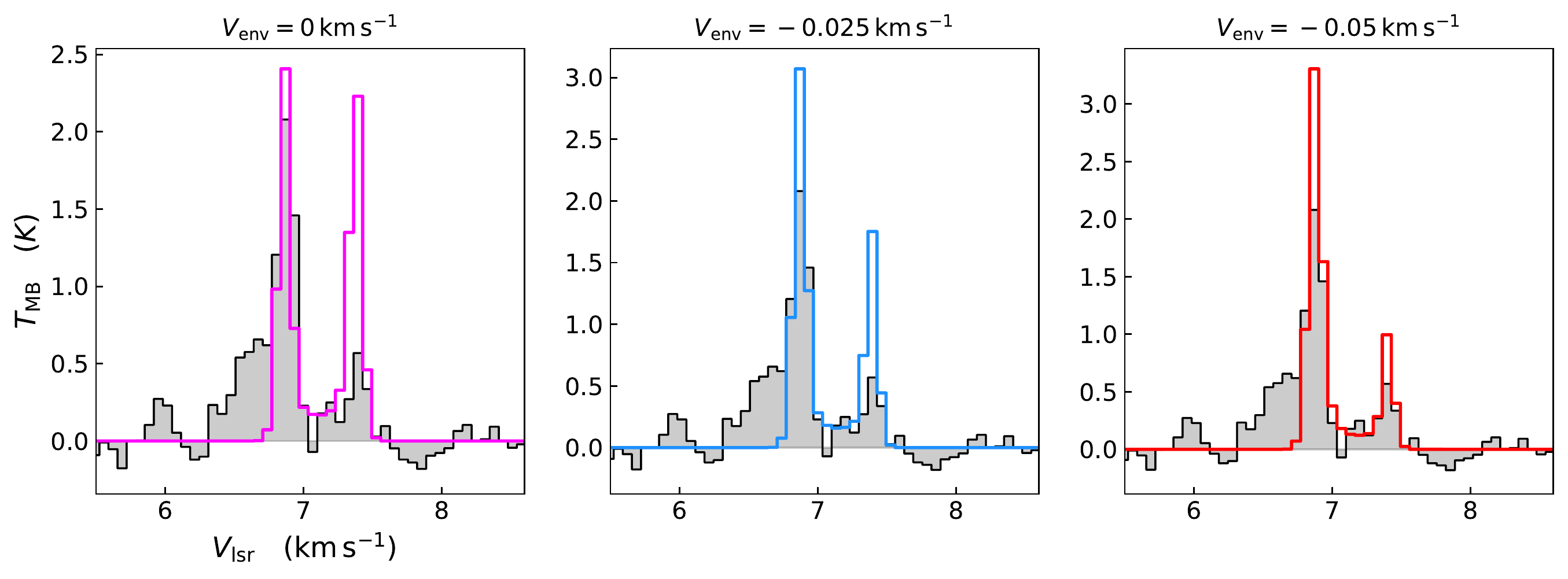}
      \caption{Synthetic spectra {of \hcop (1-0)} obtained with MOLLIE for the physical model with the extended $1.0 \rm \, pc$ envelope with a volume density of $n \rm _{env}   = 27 \, cm^{-3} $, at three distinct infall velocities, from left to right:  $V_\mathrm{env} = 0$ (static envelope, magenta histogram),  $V_\mathrm{env} = -0.025 \,$\kms (light-blue histogram), and $V_\mathrm{env} = -0.05\,$\kms (red histogram). The observed spectrum is shown with the black histogram and grey shadow.  \label{MOLLIEVelenv}}
   \end{figure*}

This last model, however, still fails in reproducing the strong blue/red asymmetry. This line profile arises from the contraction motions, and it is hence not surprising that the model with $V_\mathrm{env} = 0\,$\kms is not reproducing this feature. We hence perform two more tests, maintaining the envelope density on  $n \rm _{env}   = 27 \, cm^{-3}$, and increasing the infall velocity (in absolute value) to $V_\mathrm{env} = -0.025 \,$\kms and $V_\mathrm{env} = -0.05 \,$\kms, respectively. These tests are compared in Fig. \ref{MOLLIEVelenv}. As the absolute value of the envelope velocity increases, the blue/red peak ratio increases, from 1 to 1.5, to 3.3. This latter value is close to the observed one ($\approx 4.0$). 

\par

The last model, which is characterised by an envelope density of  $n \rm _{env}   = 27 \, cm^{-3} $ and a contraction velocity of $V_\mathrm{env} = -0.05\,$\kms, is able to reproduce the asymmetry between the blue and the red peak, and the depth and width of the central dip. It still overestimates the intensity of the two peaks, by 65\% (blue) and by 75\% (red). \par
We want to highlight that this analysis carries several limitations. First of all, the physical model presents strong uncertainties, which are aggravated by the choice of a simplistic uniform and isothermal envelope. {Furthermore, chemical simulations in general are affected by several kinds of uncertainties, for example: poorly constrained elemental abundances; unknowns in the microphysics on grain surfaces (thermal diffusion vs. tunneling, etc.); uncertainties in the reaction rate coefficients, and so on.} As a consequence, our results should be interpreted as indications of the overall physical properties of the envelope. Our analysis suggests, in conclusion, that the core L1544 is surrounded by a diffuse envelope that has a density of a few tens per cubic centimetre and is still rich in protonated carbon monoxide ($X\rm _{mol} (HCO^+) = 10^{-7}$). This envelope is not static, but it is contracting, with an inward velocity of a few $\times 10^{-2}\, $\kms. The total \hcop column density, computed from the model after convolving it to the IRAM beam size, is $N\rm _{col}(HCO^+) = 9.1 \times 10^{13} \, cm^{-2}$, which is consistent with the value  $(9.5 \pm 0.6)\times 10^{13} \, \rm cm^{-2}$ obtained by \cite{Redaelli19} using the optically thin \hcdop (1-0) line (assuming the standard isotopic ratio $\rm ^{16} O / ^{18}O = 557$, \citealt{Wilson99}). The envelope is so thin that the excitation temperature of the \hcop (1-0) line is close to the background temperature ($T_\mathrm{ex} \approx 3\, \rm K$), and therefore gas does not emit this rotational line, but it is only able to absorb the emission coming from the higher density core.
\par
    \begin{figure}[!h]
    \centering
   \includegraphics[width=.48\textwidth]{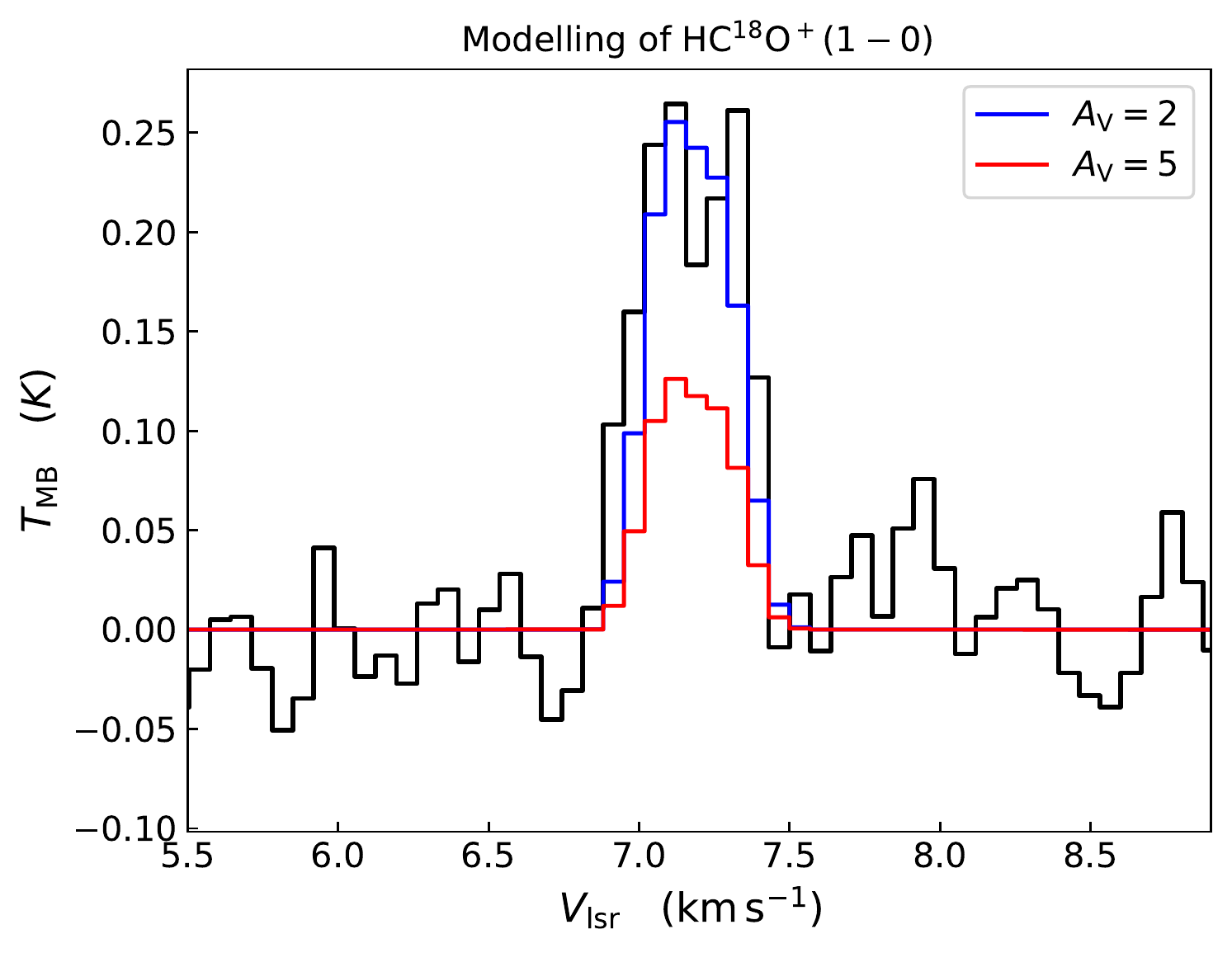}
      \caption{ The observed \hcdop (1-0) spectrum is shown with the black histogram (taken from \citealt{Redaelli19}). The blue and red histograms show the synthetic spectra obtained with MOLLIE using the chemical model with external $A_\mathrm{V} = 2$ and $A_\mathrm{V} = 5$, respectively. \label{hc18op}}
   \end{figure}
We have tested the prediction of the model also for the case of the optically thin \hcdop. As already noted in \cite{Redaelli19}, in order to reproduce this species, a chemical model with lower external visual extinction with respect to the one used to reproduce the main isotopologue is needed. The reason for this is to be found in selective photodissociation. {The main precursor of \hcop is CO, which is able to self-shield, unlike its rarer isotopologue \cdo. This can contribute to the fractionation of \hcop with respect to \hcdop in the envelope.} As a result, it is likely than the abundance of the $^{18}\rm O$-bearing isotopologue is lower than what expected taking into account only the isotopic ratio $\rm ^{16} O / ^{18}O = 557$ in the external envelope.\par
 Our new modelling confirms this hypothesis. The synthetic spectrum produced using the abundance profile derived from the chemical model with $A_\mathrm{V} = 5\, \rm mag$ underestimates the observed flux by a factor of $\approx 2$, whilst a better agreement is obtained with the abundance profile obtained adopting $A_\mathrm{V} = 2\, \rm mag$, as shown in Fig. \ref{hc18op}. The latter model matches the observed intensity, but it fails to reproduce the symmetric double-peak profile that is visible in the observations. We highlight however how these two peaks protrude above the central dip only by $\approx 80 \, \rm mK$, and that the noise of the spectrum is $rms= 38\, \rm mK$. The peaks are hence detected above a $2\sigma $ level, but the fact that they appear symmetric could be due to the limited sensitivity of our observations. We report in Appendix~\ref{hc18op_ab} the abundance profiles used to produce the synthetic spectra. \par
 The development of a chemical code which implements both the oxygen fractionation and the molecular self-shielding, which hence can self-consistently take into account the variation of the molecular abundances due for instance to photodissociation, will likely help reproducing better the \hcop isotopologues.
\par
It is worth commenting on the high external visual extinction ($A_\mathrm{V} = 5 \, \rm mag$) that we have used to increase the abundance of the main isotopologue in the external layer. The thin envelope we have manually added accounts only for a fraction of this value ($A_\mathrm{V} < 0.2 \, \rm mag$ for  $n \rm _{env}   <100  \, cm^{-3} $ ). It is likely that the remaining part of the cloud surrounding the core does not have a column density high enough to justify $A_\mathrm{V} = 5 \, \rm mag$, which furthermore would cause changing in the dust and gas temperature profiles\footnote{To these regards, we highlight that Appendix A of \cite{Redaelli21} discussed already how variations of the gas temperature of the order of $\pm 1\, \rm K$ do not affect significantly the results of the chemical modelling and radiative transfer simulations.}. However, there are other mechanisms able to increase the abundance of \hcop at low densities. Self-shielding {of the main precursor of \hcop (i.e., CO)}, can contribute in this sense. Other forms of desorption of CO from the dust grains back into the gas phase can also play a role. By using an increased external visual extinction, we simulate these processes, the modelling of which is beyond the scopes of the present work. Further details are given {in Sect. \ref{PDRmodels} and} at the end of Sect. \ref{CO_mod}.
\par
Adding the extended $1\, \rm pc$ envelope does not inficiate the results on modelling other species done in previous works \citep[see for instance][]{KetoRybicki10, Bizzocchi13, Keto15, Redaelli19, Redaelli21}. High-density tracers such as \nnhp isotopologues, or \dcop, in fact, are not expected to be abundant at the low number density that we derive in the envelope (a few tens of cubic centimetres). \hcop has been successfully detected also in the diffuse medium, unlike \nnhp \citep{Liszt94, Lucas96}. Another key difference lies in the low effective critical density of \hcop, when compared for instance to that of the \nnhp and \nndp transitions. This results in the fact that the lines of these species are not significantly affected by a change in the external visual extinction, unlike \hcop\footnote{We refere to the discussion made in Sect. 3.2.2 of \cite{Redaelli19} for further details.}.

\subsubsection{The abundance of \hcop in the envelope: comparison with diffuse clouds\label{PDRmodels}}
{The \hcop abundance in the envelope that we use is $X\rm _{mol} (HCO^+) = 10^{-7}$, which comes from extending the model at the border of the initial $0.32\, \rm pc$ up to a total radius of $1.0 \rm \, pc$. As already mentioned, this is a very simplified model, which assumes a uniform distribution of all envelope properties. Likely, the envelope has a structure (in density, temperature, and chemical composition), but its derivation would required a full hydrodynamic re-modelling of the source. This is however beyond the scope of the present work, which is to infer the general properties of the gas surrounding the dense core needed to reproduce the observations. It is nevertheless worth to compare our results, in particular in terms of \hcop abundance, with previous works and with models, to understand if they are physically meaningful.}. \par
\begin{figure}[!h]
    \centering
   \includegraphics[width=.48\textwidth]{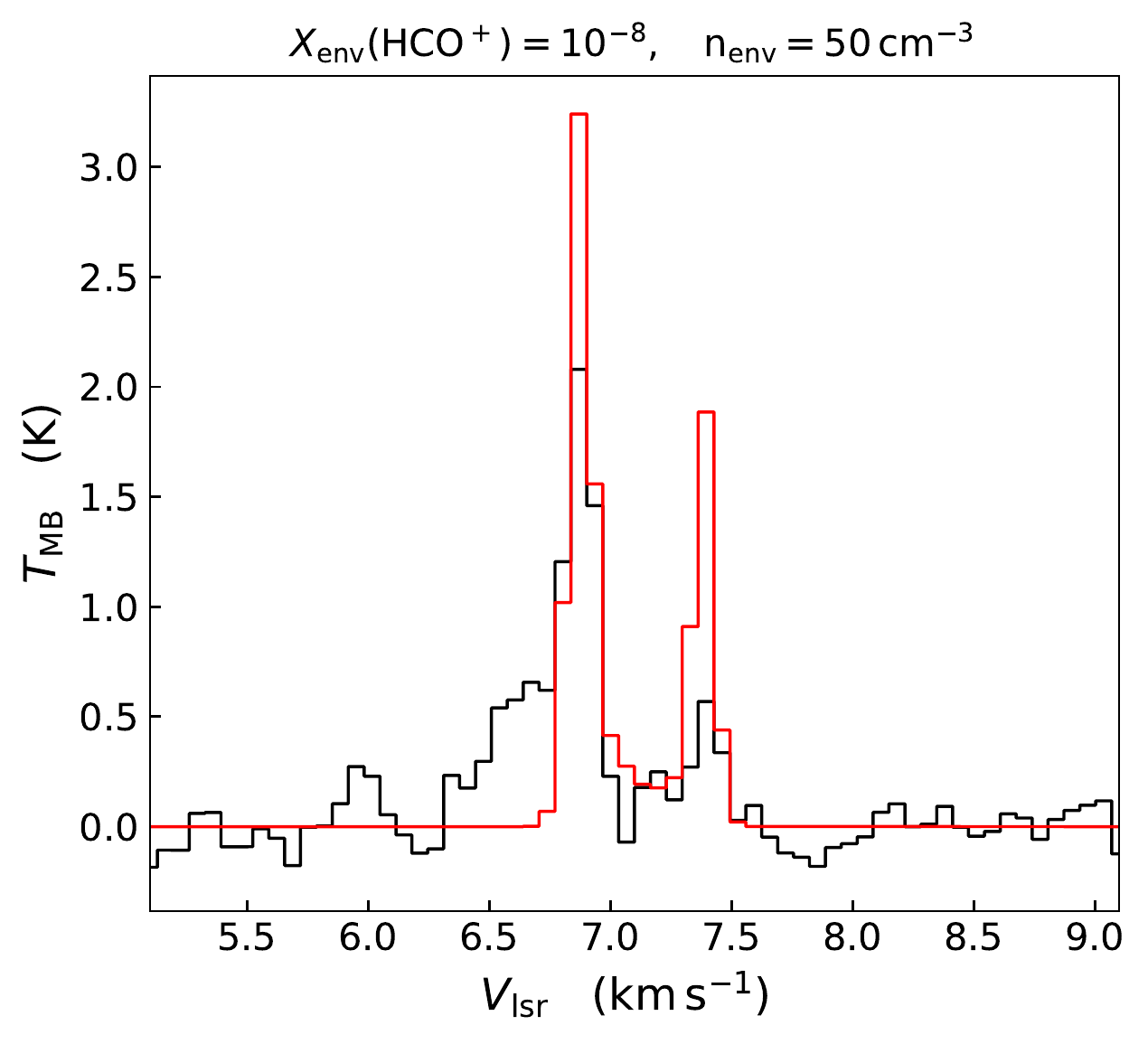}
      \caption{{ The red curve shows the synthetic spectrum of \hcop (1-0) obtained with a model with $n_\mathrm{env} = 50 \, \rm cm^{-3}$, $X_\mathrm{env} (\rm HCO^+)= 10^{-8} $, and otherwise identical to that used to produced the rightmost panel of Fig.~\ref{MOLLIEVelenv}. The black histogram shows the observed spectrum at the dust peak of L1544. }\label{hcop_test}}
   \end{figure}

{Several works have determined the abundance of \hcop in diffuse clouds, using absorption spectroscopy towards bright background sources. The usual abundance values found are of the order of $X(\rm HCO^+) = 0.1-1.0 \times 10^{-8}$ \citep[see e.g.][]{Lucas93, Lucas96, Gerin19}, i.e. at least an order of magnitude lower than our estimation. This does not come as a surprise; it is likely that the abundance of \hcop has indeed a decreasing gradient in the envelope, and that therefore its average abundance is lower than $X(\rm HCO^+) = 10^{-7}$ found at $r=0.32\, \rm pc$. To understand whether a lower abundance would still reproduce our observations, we have made a test, running the radiative transfer with the following envelope properties: $n_\mathrm{env} = 50 \, \rm cm^{-3}$, $X_\mathrm{env} (\rm HCO^+)= 10^{-8} $ ; the other physical properties are identical to those used to obtain the best agreement with the observed \hcop spectra in the previous subsection. The synthetic spectrum obtained with this model, compared with the observations, is shown in Fig.~\ref{hcop_test}. Overall, the agreement with the observations is still acceptable (in particular, in the level of the flux dip due to self-absorption), even though the blue/red asymmetry is less strong. The \hcop column density obtained with this model, furthermore, is only $14$\% lower than what obtained using \hcdop observation \citep{Redaelli19}.   Given the uncertainties and the simplistic assumptions of our model, we conclude that the abundance of \hcop in the envelope needed to reproduce the observations is $10^{-8}-10^{-7}$, and that this value is consistent in its lower limit to observational results in the diffuse medium.} \par

   {The diffuse envelope resembles a photodissociation region (PDR), where the chemistry is dominated by the impinging interstellar UV flux, and we have searched for a comparison in the Meudon database for PDR models \citep{LePetit06}, available online (\url{https://app.ism.obspm.fr/ismdb/}). The isobaric model with thermal pressure $P = 10^3 \, \rm cm^{-3} \, K$, maximum visual extinction $A_\mathrm{V} = 2 \, \rm mag$, and illuminated by the standard interstellar radiation field \citep{Mathis83}, is characterised by temperature and density values close to those used in our envelope model in their order of magnitude ($T \approx 70 \, \rm K $, $n \approx 10 \, \rm cm^{-3}$). The predicted peak abundance is $X (\rm HCO^+)= 10^{-9} $, at least one order of magnitude lower than what our results suggest. This point has however been explored in detail by \cite{Godard10}, who performed a thorough comparison between the observations of \hcop (and other molecular tracers, e.g., cyanides) in the diffuse medium and two kinds of chemical models: \textit{i)} the Meudon PDR models; and \textit{ii)} the turbulent dissipation region (TDR) model \citep{Godard09}. Those authors found that the PDR models underestimate the observed \hcop column densities by one order of magnitude, or more. On the contrary, the TDR models are well in agreement with the observational data, as they predict higher \hcop abundances \citep[$10^{-8}-10^{-7}$, see Fig. 3 of][]{Godard09}. More recently, \cite{Rybarczyk22} also noted anomalous large column densities of \hcop in diffuse clouds, which might be due to non-equilibrium chemistry. There are other effects, furthermore, that can enhance the \hcop abundance with respect to what is predicted by PDR models. For instance,  \cite{CecchiPestellini00} suggested the existence of small (au in size), dense cloudlets where molecules can form efficiently, to then evaporate in the diffuse surrounding medium. The existence of such dense cloudlets has been theoretically confirmed by \cite{Tsytovich14}. }  
   
\subsubsection{\co modelling \label{CO_mod}}
   \begin{figure}[!h]
    \centering
   \includegraphics[width=.5\textwidth]{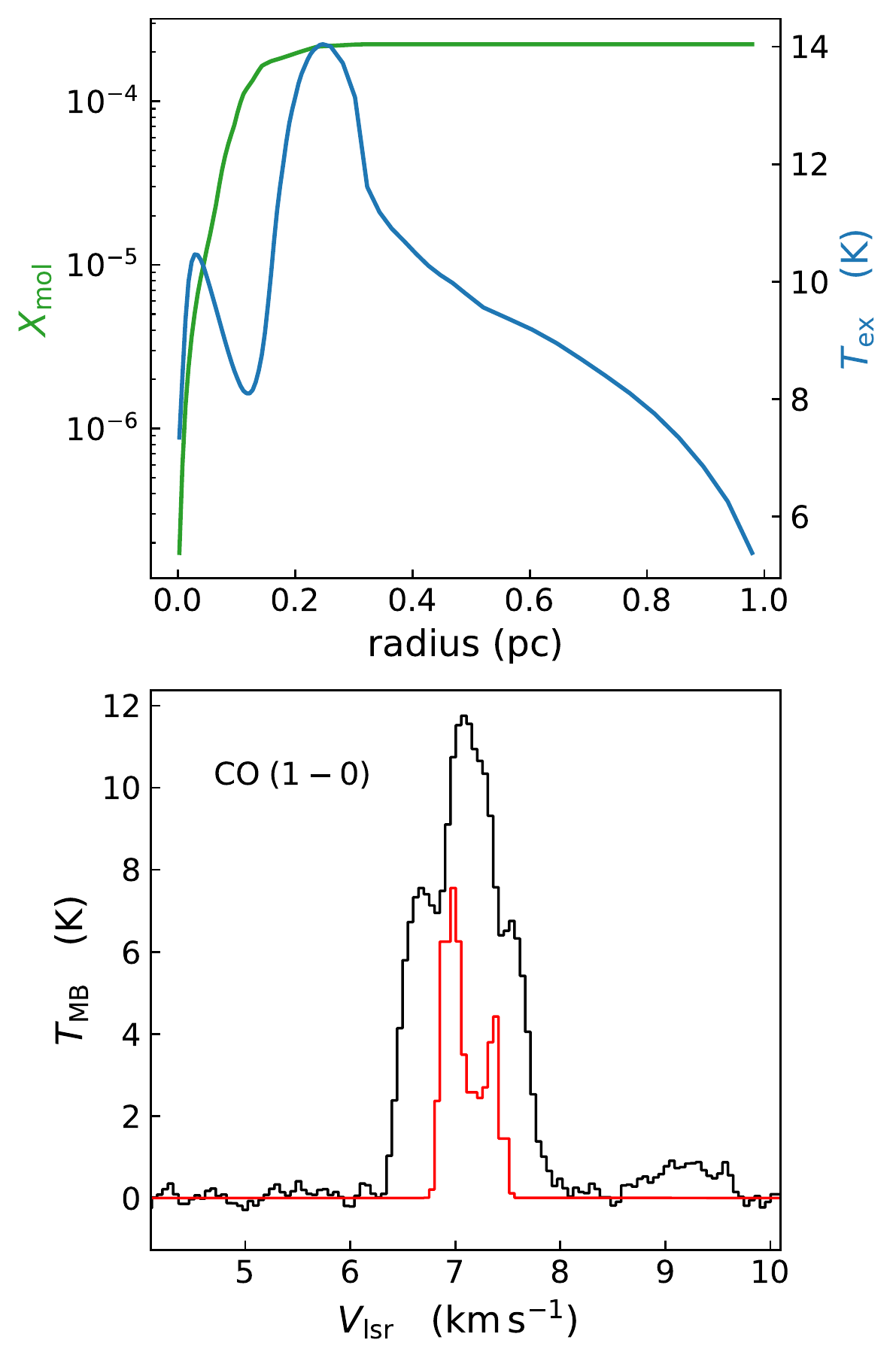}
      \caption{ \textit{Top panel:} The green curve shows the $\rm CO$ abundance profile predicted by the chemical model at $t=10^6 \, \rm yr$ and with an external visual extinction $A_\mathrm{V} = 5\, \rm mag$, extended in the $1\, \rm pc$ envelope. The blue curve shows the excitation temperature profile of the (1-0) transition computed by MOLLIE.  \textit{Bottom panel:} the observed $\rm CO$ (1-0) spectra towards the L1544 dust peak (black histogram) compared with the synthetic spectra computed with MOLLIE using the abundance profile shown in the top panel (red histogram). The observed spectra is from Chac\'on-Tanarro et al. (in prep). \label{MOLLIEco}}
   \end{figure}
The main formation pathway of \hcop in molecular gas is from carbon monoxide (CO). It is hence important to verify how the best model found for \hcop behaves in reproducing the {precursor of this species}. To this aim, we use the CO (1-0) spectrum at the dust peak of L1544 observed with IRAM (details about these data will be published in Chac\'on-Tanarro et al., in prep). We produced the synthetic spectrum with MOLLIE adopting the physical model that provides the best agreement for the \hcop case. The abundance profile adopted for this simulation is shown in the top panel of Fig. \ref{MOLLIEco}, and it is derived from the chemical code in the same conditions as for the \hcop one.  \par
The CO (1-0) spectrum is significantly more complex than the \hcop one, due to the presence of several velocity components. Beyond a broad and well separated feature at $\approx 9\, $\kms, likely arising from a more diffuse cloud on the line of sight {\citep[this spectral feature has been already detected in the FCRAO survey of Taurus;][]{ Goldsmith08,Narayanan08}}, the main component appears split into three peaks: the brightest one is centred at L1554 rest velocity. Two, fainter ones are visible on both blue and red sides, with peak intensities of $6.5-7.5\, \rm K$, and separated in velocity of $\approx 0.5 \, $\kms. The origin of these two components remains unknown. We could speculate that they represent the original filaments that collided to form the L1544 core and the dense filament within which is embedded. \par
The MOLLIE spectrum, shown in red in Fig. \ref{MOLLIEco} (bottom panel), does not reproduce the observed data. It underestimates severely the observed flux (by a factor of $\approx 2$), and it present a strong self-absorption feature, with blue-peak asymmetry, similar to the one observed in the \hcop line. We performed the same test on \cdo (see Sect. \ref{cdo_model} for more details). Also in this case, the simulated lines are underestimated with respect to the observed ones.  \par
The fact that we are not able to reproduce correctly carbon monoxide isotopologues can be due to different reasons. In particular, in the diffuse envelope, where the density is only a few tens of $ \rm cm^{-3}$, different reactions can be dominant, {and it is hence possible that the \hcop/CO abundance ratio is different to that predicted by our chemical code at higher densities ($n > 10^2 \rm \, cm^{-3}$).}

\subsection{Analysis of the map with the Hill5 model}
   \begin{figure*}[!t]
    \centering
   \includegraphics[width=\textwidth]{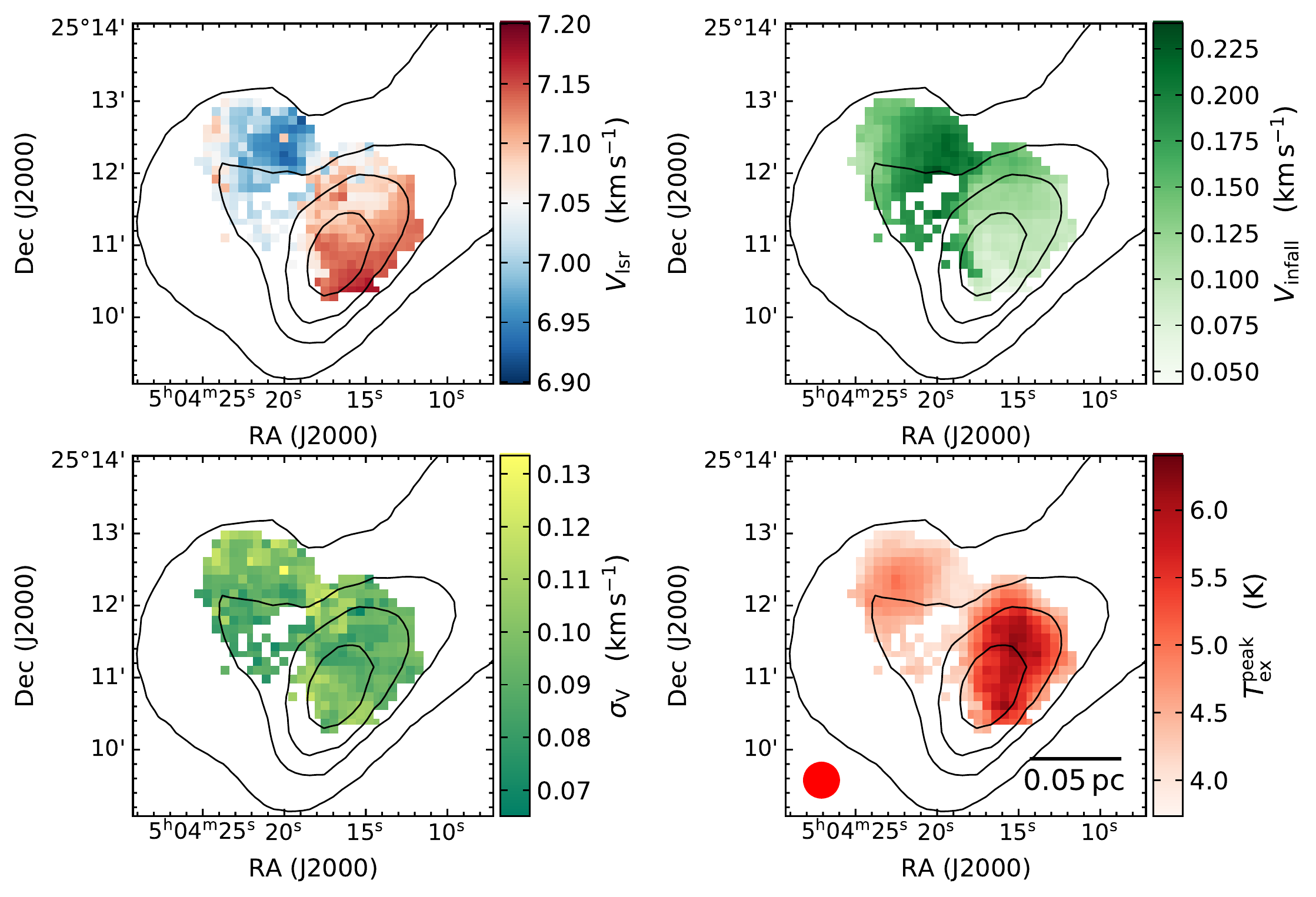}
      \caption{Maps of the best-fit values obtained with the two-step fitting procedure with the Hill5 model implemented in \textsc{pyspeckit}. The panels refer to \vlsr (top-left), \vin (top-right), \sigmav (bottom left), and $T\rm _{ex} ^{peak}$ (bottom-right). The solid contours show the H$_2$ column density, as in Fig. \ref{mom0} (note however that the maps have been zoomed-in). In the bottom-right panel the beam size and scalebar are also shown.  \label{Hill5}}
   \end{figure*}
The non-LTE analysis performed with MOLLIE is limited to the line-of-sight going through the dust peak, for which the physical model has been developed. In order to investigate the properties of the contracting envelope offset from the core's centre, we make use of the so-called Hill5 model \citep{deVries05}, implemented in \textsc{pyspeckit}. It represents an improvement with respect to the two-layer model used to reproduce strongly absorbed line profiles for instance by \cite{Tafalla98}, since it is likely a better representation of real cores compared to a two-slabs geometry. In the Hill5 model, the excitation temperature of the analysed transition is assumed to increase linearly with the optical depth, and it increases from a minimum value at the source edge $T\rm _{ex} ^0 $ to the peak value $T\rm _{ex} ^{peak}$ at the centre. The core contracts with a constant infall velocity $V_\mathrm{infall}$. In the  \textsc{pyspeckit} implementation, $T_\mathrm{ex} ^0  = T_\mathrm{bg} = 2.73 \, \rm K$. Hence, the free parameters of the model are: the peak value of the line optical depth $\tau \rm  ^{peak}$,  the line centroid velocity \vlsr, the infall velocity with respect to \vlsr ($V_\mathrm{infall}$), the line velocity dispersion $\sigma_\mathrm{V}$, and $T\rm _{ex} ^{peak}$. {By applying it pixel-per-pixel in the whole map, we are probing the spatial variation of the line-of-sight component of the infall velocity (which as a whole is directed toward the central part of the core).} \par
\cite{deVries05} suggest to use the model only on high signal-to-noise ratio data, since given the high number of free parameters, the fitting procedure requires high sensitivity data to converge. We hence first mask all pixels with $\rm S/N< 15$ (in peak temperature; this leaves 357 pixels). In order to improve the fit convergence, we limit the free parameter space using the following conditions:
$\tau \rm  ^{peak} \in [0;100] $, $ V_\mathrm{lsr} \in [6;8] \, $\kms, $\sigma_\mathrm{V} \in [0;0.2] \, $\kms; $T\rm _{ex} ^{peak}$ and  $V_\mathrm{infall}$ are limited to positive values. Note that in the Hill5 implementation $V_\mathrm{infall}> 0$ means infall motions, whilst in the physical model used in MOLLIE the opposite convention holds. \par
After this first fitting procedure, for a large number of pixels (65 \% of the positions that satisfy $\rm S/N>15$) the best-fit value of the optical depth hits the upper limit $\tau \rm  ^{peak} = 100 $, or the uncertainty on best-fit value for $\tau^\mathrm{peak}$ is higher than 50\%. This happens in particular towards the north-east direction, where the red wing of the spectral profile is absorbed almost to the zero flux level. In this situation, the optical depth is so high that it becomes a degenerate parameter, and the fit does not converge properly. In order to obtain a converged fit, we re-fit these pixels fixing the optical depth of the average value  $\langle \tau \rm  ^{peak} \rangle = 43$, computed on those positions where the fit converged the first time. After the second procedure, the best-fit value maps are combined with the ones obtained in the first fit. As for an example, we report a few spectra together with their best-fit in Appendix~\ref{app:Hill5}.\par
We investigate the effects of the choice of the fixed value for \taup in the second fitting procedure by repeating it in two tests, using first $\tau \rm  ^{peak} = \text{fixed} = 25$, and then $\tau \rm  ^{peak} = \text{fixed} = 75$. The changes in the best-fit values for \vlsr,  \vin, and $T\rm _{ex} ^{peak}$ are within 1\%. The results of the velocity dispersion are instead affected by $\approx 7$\%. Since our focus is on the infall velocity, we conclude that the choice of \taup does not affect significantly our results, and we hence consider the best-fit results those obtained with the double-fitting approach, using $\tau \rm  ^{peak} = \text{fixed} = 43$ in the second iteration.  The resulting maps for \vlsr, \vin, $\sigma_\mathrm{V}$, and $T\rm _{ex} ^{peak}$ are shown in Fig. \ref{Hill5}.\par
   
We derive the map of the fit residuals by computing the standard deviation of the difference between observed and modelled spectra pixel-per-pixel, in the velocity range $[6-8]\,$\kms. The mean residual is $130 \rm \, mK$, which has to be compared with the observation $rms$ ($100 \, \rm mK$), and are well below the $3 \sigma$ value, proving the good quality of the obtained fits. Higher residual values ($rms \approx 200 \rm \, mK$) are found towards the southern part of the core, where the \hcop line appear to have a broader shoulder in the blue peak (see Fig. \ref{SpectraGrid}), which cannot be reproduced by the Hill5 model alone.

\par
\begin{figure}[h]
    \centering
   \includegraphics[width=0.48\textwidth]{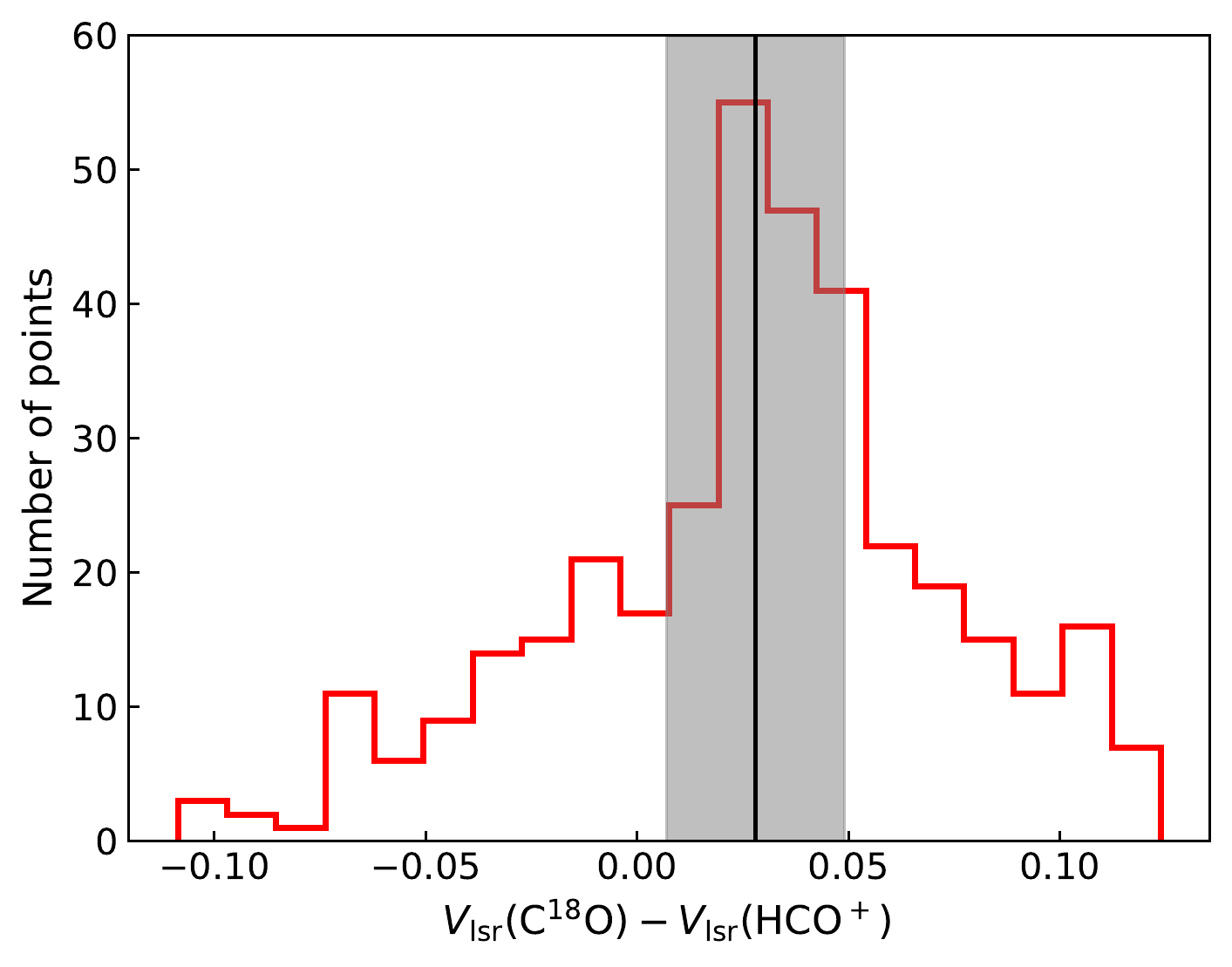}
      \caption{Histogram of the distribution of the difference between the centroid velocity of \cdo and \hcop, as obtained with the single gaussian and the Hill5 fit, respectively, in \kms. The vertical solid line is the mean of the distribution, whilst the shaded grey area shows the $3\sigma$ interval around it.   \label{VelDiff}}
   \end{figure}

As a further test, we compared the centroid velocity maps obtained from the single Gaussian fitting to the \cdo data and from the Hill 5 analysis on the \hcop data, by computing the difference $ V_\mathrm{lsr}(\text{\cdo})-V_\mathrm{lsr}(\text{\hcop})$. The distribution of this quantity is shown in Fig. \ref{VelDiff}, and its mean value is $0.028\,$\kms. The associated error, computed by adding in quadrature the uncertainties obtained from the fit procedure on the \vlsr of the two tracers, is $0.007\,$\kms. The difference $ V_\mathrm{lsr}(\text{\cdo})-V_\mathrm{lsr}(\text{\hcop})$ is hence marginally inconsistent with 0.0 at the $3\sigma$ level, even though one has to take into account the limited spectral resolution of the \hcop data ($0.065\,$\kms). However, small kinematic changes in the gas traced by these two molecules are expected, since likely \cdo, being a rarer isotopologue, is more affected by selective photodissociation in the external parts of the source (see also the discussion made in Sect.~\ref{CO_mod}).  

\par
The excitation temperature map follows the morphology of the integrated intensity map shown in Fig. \ref{mom0}, with two separated peaks, a stronger one located north-west of the core dust peak, and a second smaller one towards north-east. The peak values is $T\rm _{ex} ^{peak} = (6.24 \pm 0.11) \, \rm K$. Note that this value cannot be compared directly with the peak one obtained with MOLLIE (left panel of Fig. \ref{MOLLIEstaticenv}), since the latter is derived as a radial profile from the physical model, whilst the former represents an average both on the line-of-sight and on the beam area. It is therefore expected that the  $T\rm _{ex} ^{peak}$ presents lower values than the excitation temperature obtained from MOLLIE, taking also into account the intrinsic differences of the two methods. 
\par
The velocity dispersion map appears fairly uniform, with an average value of $<\sigma_\mathrm{V} >= 0.10 \, $\kms. This value is significantly lower than the average value obtained from the \cdo (2-1) data ($< \sigma_\mathrm{V} > = 0.16 \, $\kms). This is explained by the fact that the Hill5 model takes into account also the kinematics (i.e. contraction) of the core, which contributes to increase the linewidths obtained from a single-component gaussian fit. The value obtained from the Hill5 model does not contain the contribution from the ordered and coherent infall motions, but only the thermal and turbulent components. The non-LTE modelling requires the turbulent linewidth contribution to be introduced manually. We adopt the value $\sigma_\mathrm{turb} = 0.075\, $\kms, which allows {to} reproduce the observed linewidths (see for instance \citealt{Bizzocchi13, Redaelli18}). The thermal broadening of \hcop at $10\, \rm K$ is $0.05\, $\kms. If we sum these two components in quadrature we obtain $0.09\,$\kms, which is very similar to the $<\sigma_\mathrm{V} >= 0.10 \, $\kms value obtained in the Hill5 analysis.
\par
We finally comment the infall velocity map. The region surrounding the dust peak is characterised by $\text{\vin} \approx 0.11\,$\kms. The value at the dust peak is $\text{\vin} = (0.083 \pm 0.006)\,$\kms, which is in between the peak value of the infall velocity profile of the physical model ($0.15\, $\kms) and the envelope value  $|V_\mathrm{env}| = 0.05\,$\kms found in Sect. \ref{MOLLIEhcop}. The infall velocity then increases in the north-west portion of the core, reaching values of $0.17-0.23\, \rm$\kms. \par 
The Hill5 modelling shows that the whole envelope of L1544 traced by \hcop is contracting. This is consistent with the recent discovery that the envelope of L1544 is magnetically supercritical, as unveiled by the observations of the Zeeman effect in the HI absorption line done with the FAST telescope \citep{Ching22}. In that paper, the authors report the detection of the Zeeman splitting at a position $\approx 0.15\, \rm pc$ away from the dust peak towards the north-west direction, and they derive a value of magnetic field of $B=3.8\, \mu \rm G$. The estimated H$_2$ column density at that position is $3.5 \times 10^{21} \, \rm cm^{-2}$, and hence the mass-to-flux ratio is $\lambda = 3.5\pm 0.3$, significantly higher than the critical value $\lambda = 1.0$. If this envelope is supercritical, the gravitational force might have already led to a contraction. 

\section{Conclusions\label{conclusion}}
In this work, we presented a map of the \hcop (1-0) transition towards the prestellar core L1544, observed with the IRAM 30m telescope at a resolution of $0.024\, \rm pc$. The map footprint is large enough to comprise all positions with gas column density higher than $4\times 10^{21} \,\rm cm^{-2}$. \hcop, which is mainly formed from CO, is strongly affected by depletion at high densities. According to our chemical model, its abundance drops by more than two orders of magnitude when $n \gtrsim 10^4\, \rm cm^{-3}$.  At lower densities, the molecule has instead a higher abundance ($X_\mathrm{mol} = 10^{-7}-10^{-8}$). Its ground-state rotational transition is hence optically thick, and moreover it has a relatively large critical density ($n_\mathrm{c} = 7 \times 10^{4}\, \rm cm^{-3}$), but a low effective critical density because of the large opacities. This line is therefore an ideal probe of infall motions in the lower density envelope surrounding the core. \par
The acquired spectra show a clear double-peak profile in the whole map coverage. The two peaks often show blue asymmetry. The fainter, red peak approaches the zero flux level towards the north-west part of the source. The central flux dip is found to correspond to the centroid velocity of the optically thin \cdo (2-1) line, confirming that the \hcop line profiles arise from a combination of strong self-absorption and contraction motions along the line of sight. The data hence show that the whole envelope is experiencing gravitational contraction.  \par
We have performed a detailed modelling of the spectrum at the dust peak using a non-LTE approach. We have used our state-of-the-art chemical code to predict the molecular abundance in the physical structure modelled by \cite{Keto15}, and then the non-LTE radiative transfer code MOLLIE has been used to produce synthetic spectra. The original model, which extended up to $0.32\,\rm pc$, fails in reproducing two key features of the observed spectral profile: the flux of the central dip that reaches the zero level, and the strong blue/red peak asymmetry (with an intensity ratio of $\approx 4.0$). In order to reproduce the observed features, a stronger absorption is needed. We have hence simulated a uniform, isothermal envelope that extends to $1.0 \rm \, pc$. We found that in order to obtain a good agreement with the observations, the envelope must be low density (a few tens of $ \rm cm^{-3}$), and it cannot be static, but requires an infall velocity. The best-fit model is found for $n _\mathrm{env} = 27 \, \rm cm^{-3}$ and $V_\mathrm{env} = -0.05 \, $\kms. Given the limitations of our analysis, we interpret these results not as exact numbers, but as an indication of the average properties of the envelope. This envelope is so diffuse that the \hcop does not emit in rotational lines, but instead it absorbs the bright emission coming from the central denser core. Indeed, unlike other high-density tracers such as \nnhp, \nndp, and \dcop, \hcop has been successfully detected in the diffuse medium \citep{Lucas96}. The presence of such low-density, large scale gas is likely common to other star-forming regions. For instance, \cite{Pineda08} found that larger fraction of the CO emission is subthermally excited, i.e. it must come from densities lower than the critical density of the (1-0) transition ($n_\mathrm{crit}\approx 10^3 \rm \, cm^{-3}$).
\par
Concerning the size of the envelope, it is worth mentioning that \cite{Li03} observed the $ \rm H\,{\textsc i}$ absorption transition at a position $\approx 0.15\, \rm pc$ away from the dust peak of L1544, and they derived a column density of $N(\rm H\,{\textsc i}) = 4.8\times 10^{18} \,  cm^{-2}$. According to calculations made by \cite{Goldsmith05} concerning the volume density of \hbox{\rmfamily H\,{\textsc i}}, both chemical calculations and observational evidence suggest that its volume density is fairly independent of the total gas density, and it is of the order of $n(\rm H\,{\textsc i})= 2\,  cm^{-3}$. Taking this into account, the length along the line of sight probed by the Zeeman observations in \cite{Ching22} is $\approx 0.8 \, \rm pc$. This value is of the same order of the size of the simulated envelope we discussed in Sect. \ref{MOLLIEhcop}. 
\par
The non-LTE analysis is limited to the dust peak position, since we do not have a fully three-dimensional model of the core. In order to model also the spectral cube data, we have used the Hill5 model developed by \cite{deVries05}. The obtained \vin map confirms that contraction motions with velocities $0.1-0.2\,$\kms are detected at all the positions where we perform the fit. The average velocity dispersion of the line, $0.1\,$\kms, can be explained as a sum of the contribution of thermal broadening ($0.05\,$\kms) and turbulent motions ($\sigma_\mathrm{turb} = 0.075\,$\kms according to the non-LTE analysis). \par
Our results show that the contraction motions in L1544 are not limited to the dense core, but they extend well in the envelope. This is difficult to explain in the inside-out collapse scenario proposed by \cite{Shu77}, and it is apparently more consistent with the outside-in model of the Larson-Peston flows, which however predicts supersonic motions towards the edge of the core, and it does not reproduce the line profiles of high density gas tracers (see \citealt{Keto15}) \par
Our results are consistent with the recent measurement of the magnetic field in the L1544 envelope, based on the Zeeman splitting of the HI absorption line \citep{Ching22}. According to those authors, the envelope is magnetically supercritical, and the gravitational force is enough to overcome the resistance of the magnetic field already at low column densities ($\approx 4 \times 10^{21} \rm cm^{-3}$). The contraction might also be triggered by the collisions of two filaments or filamentary-likes structures, a scenario which would also explain the multiple components seen in the CO line profile.\par
Our work gives a first insight on the kinematics of the gas surrounding the dense core. As future perspectives, the development of a fully 3D model of the source, which can then be coupled with radiative transfer analysis also at different offsets from the dust peak, will provide more information to constrain the properties of the contracting envelope. Such a model should be coupled with time-dependent chemistry evolution, and possibly including also molecular self-shielding, in order to reproduce simoultaneously the carbon monoxide and the \hcop isotopologues.

\restartappendixnumbering

   \begin{acknowledgements}
ER acknowledges the support from the Minerva Fast Track Program of the Max Planck Society. ER, {SS}, and PC acknowledge the support of the Max Planck Society.  AC-T and MT acknowledge partial support from project PID2019-108765GB-I00 funded by  MCIN/ AEI /10.13039/501100011033.
  \end{acknowledgements}

\software{{GILDAS \citep{Pety05,GILDAS13}, \textsc{pyspeckit} package \citep{Ginsburg11}, MOLLIE \citep{Keto90,Keto04}, Meudon PDR model \citep{LePetit06}}}
\newpage
\appendix

\section{Unscaled \hcop (1-0) spectra\label{UnscaledSpectra}}
\begin{figure*}[!b]
    \centering
   \includegraphics[angle = -90 , width=.85\textwidth]{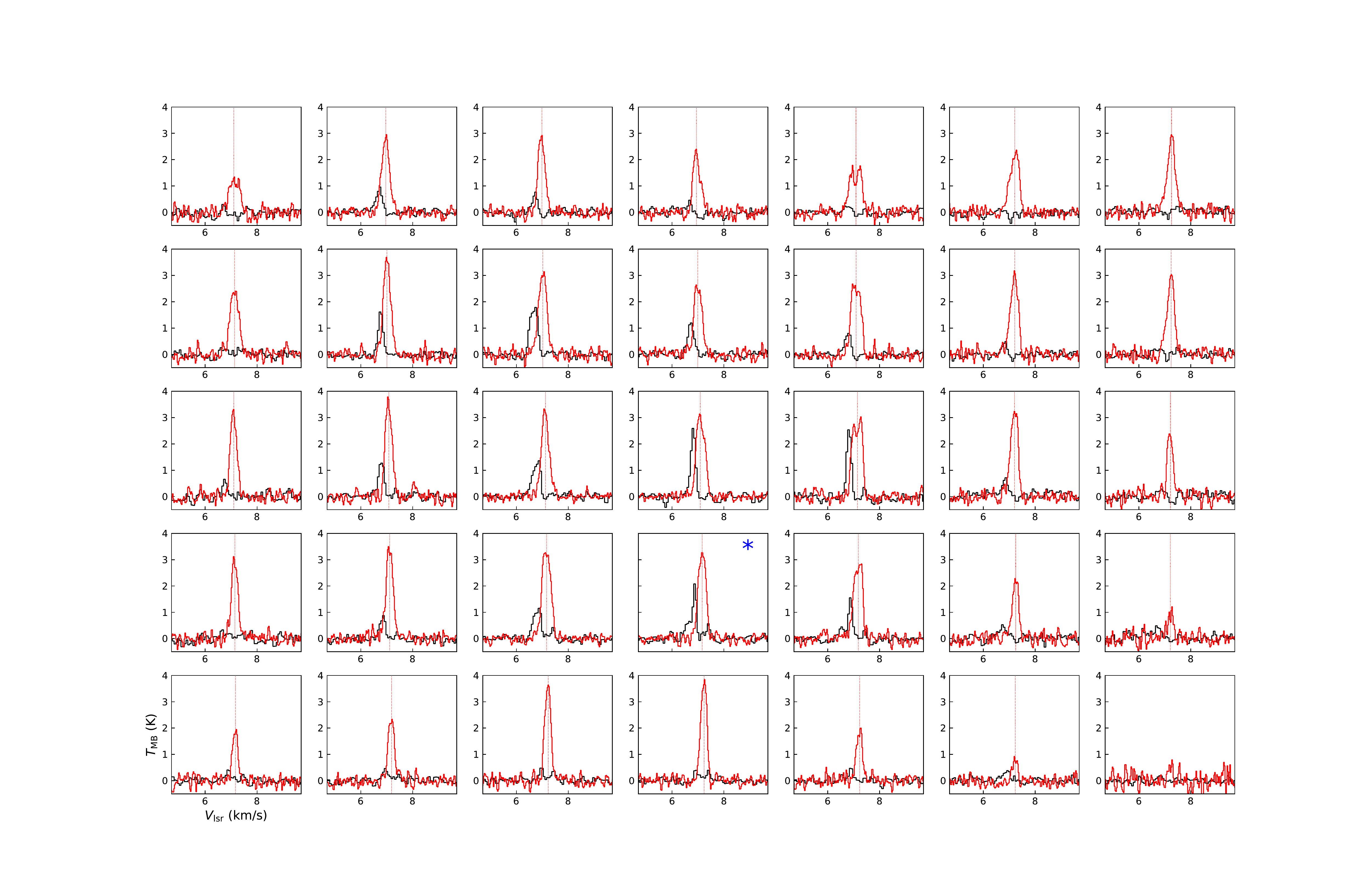}
      \caption{{Black histograms are the \hcop spectra at the positions shown with black crosses in Fig. \ref{mom0}, whilst red histograms show the corresponding \cdo (2-1) spectra. The panel with the blue asterisk represents the spectrum at millimetre dust peak. The data are the same shown in Fig.~\ref{SpectraGrid}, but this time the intensity axis is unscaled and shows the observed values.}\label{UnscaledFig}}
   \end{figure*}

{Figure~\ref{UnscaledFig} reports the same spectra of \hcop (1-0) and \dcop (2-1) lines shown in Fig.~\ref{SpectraGrid}, but in observed intensity units. The y-axis range is kept fixed in all panels, to better appreciate the intensity variation throughout the map footprint. }

\clearpage
\section{\hcdop abundance profiles\label{hc18op_ab}}
\begin{figure}[h]
    \centering
   \includegraphics[width=.5\textwidth]{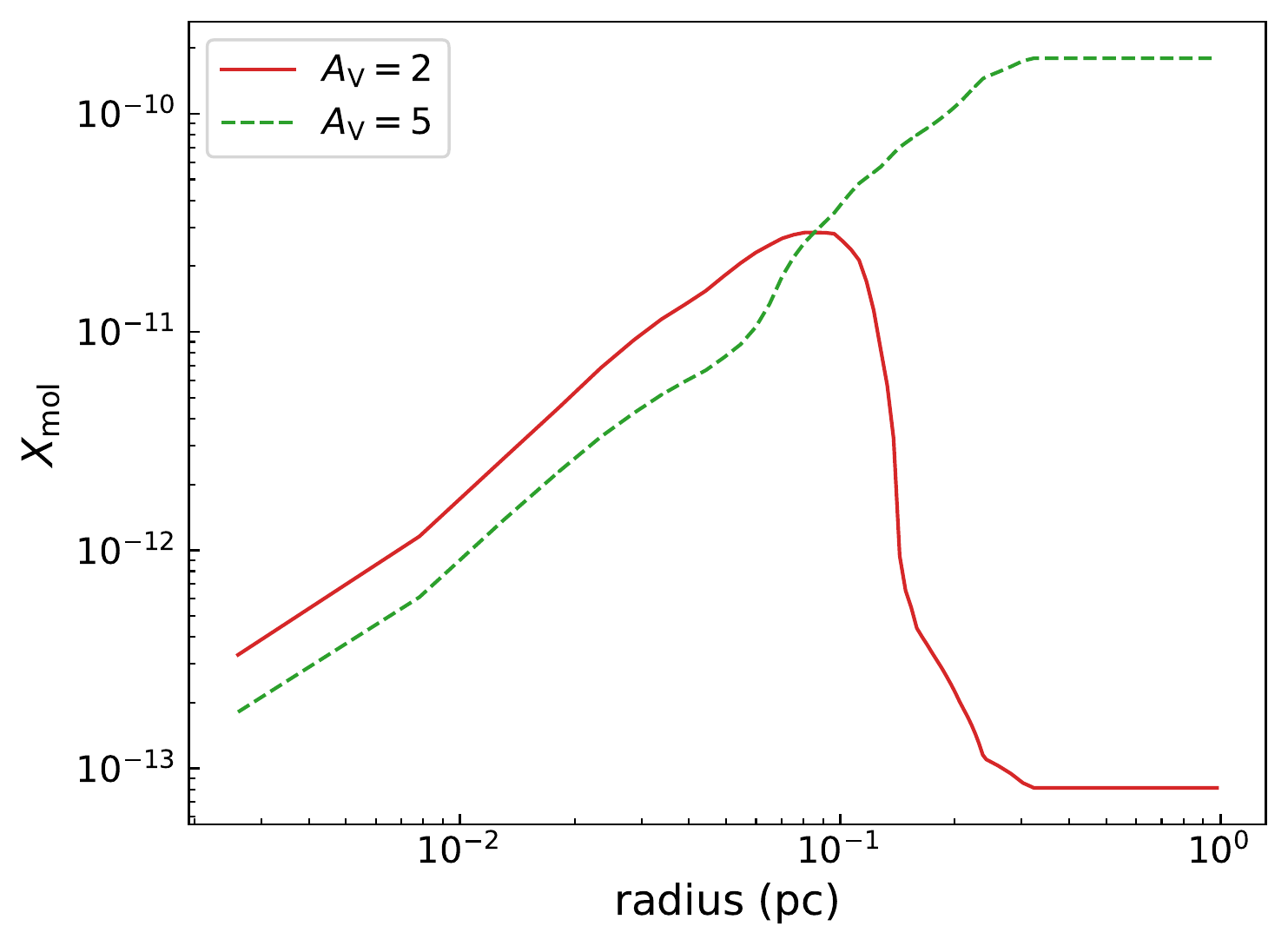}
      \caption{ Abundance profiles obtained for \hcdop with our chemical code, at $t = 1\, \rm Myr$. The red, solid curve shows the model where a lower external visual extinction is implemented ($A_\mathrm{V} =2  \, \rm mag$), whilst the higher value  $A_\mathrm{V} =5 \, \rm mag$  is used to produce the dashed green line. The two profiles have been extended in the simulated envelope. \label{hc18op_ab_fig}}
   \end{figure}

We report in Fig. \ref{hc18op_ab_fig} the abundance profiles of \hcdop derived from the chemical code using $A_\mathrm{V} =2  \, \rm mag $ and  $A_\mathrm{V} =5 \, \rm mag$. The profiles have been extended in the envelope, assuming constant molecular abundance, as done for \hcop. Since the chemical code does not implement the fractionation of oxygen, the $X \rm _{mol} (HC^{18}O^+)$ profiles have been derived from the corresponding ones of the main isotopologues, using the standard isotopic ratio $\rm ^{16} O / ^{18}O = 557$. Using two different external extinction values to reproduce the two isotopologues can be explained considering that \hcdop, being much rare, is not able to self-shield, and it is hence more affected by photodissociation in the external parts of the source.
   
         \begin{figure*}[!b]
    \centering
   \includegraphics[width=.6\textwidth]{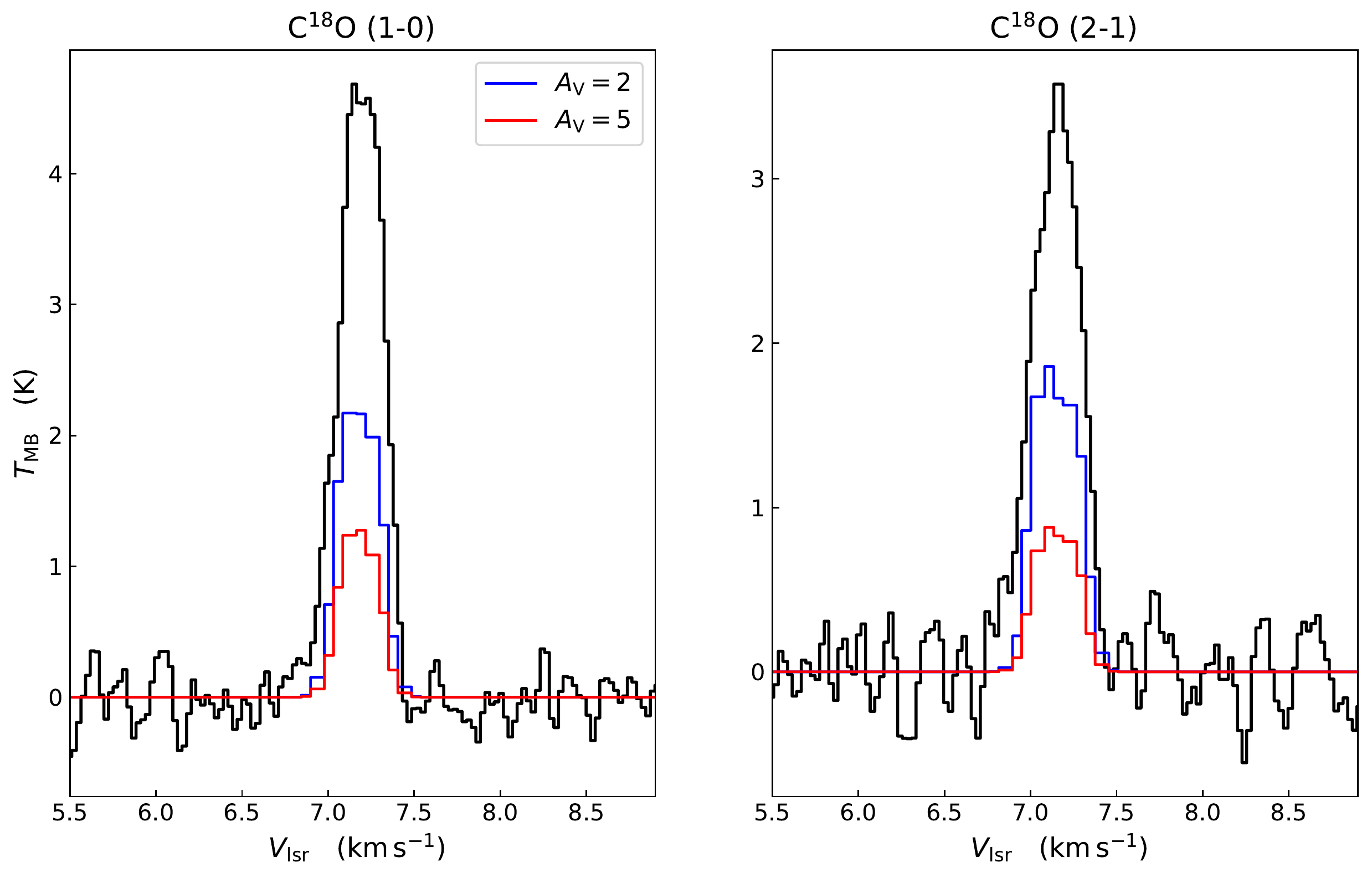}
      \caption{\textit{Left panel}: the observed (1-0) spectrum of \cdo is shown in black, overlaid to the synthetic line profiles obtained with MOLLIE using different values for the external visual extinction: $A_\mathrm{V} =2\, \rm mag $ and  $A_\mathrm{V} =5\, \rm mag$ are shown in blue and red, respectively. \textit{Right panel:} same as in the left one, but for the (2-1) line. \label{c18o_MOLLIE}}
   \end{figure*}
   
   \section{MOLLIE modelling of \cdo \label{cdo_model}}
   In Fig. \ref{c18o_MOLLIE} we show the observed \cdo (1-0) and (2-1) spectra observed with the IRAM 30m telescope at the dust peak of L1544 (from Chac\'on-Tanarro et al., in prep). The corresponding synthetic spectra obtained with MOLLIE using the chemical model with external extinction of $A_\mathrm{V} =2 \, \rm mag$ and  $A_\mathrm{V} =5\, \rm mag$ are shown in blue and red, respectively. Similarly to \hcdop, using a lower extinction value improve the agreement, but the peak intensity are still underestimated by a factor of 2. This suggests that similarly to protonated carbon monoxide, also for CO isotopologues fractionation and/or selective photodissociation may play a role. At the same time, the fact that we cannot reproduce these observations highlights how a self-consistent physical and chemical model of the extended envelope is needed, in order to fully characterise its properties.

\section{Selection of spectra fit with Hill5 model\label{app:Hill5}}
\begin{figure*}[h]
    \centering
   \includegraphics[width=.9\textwidth]{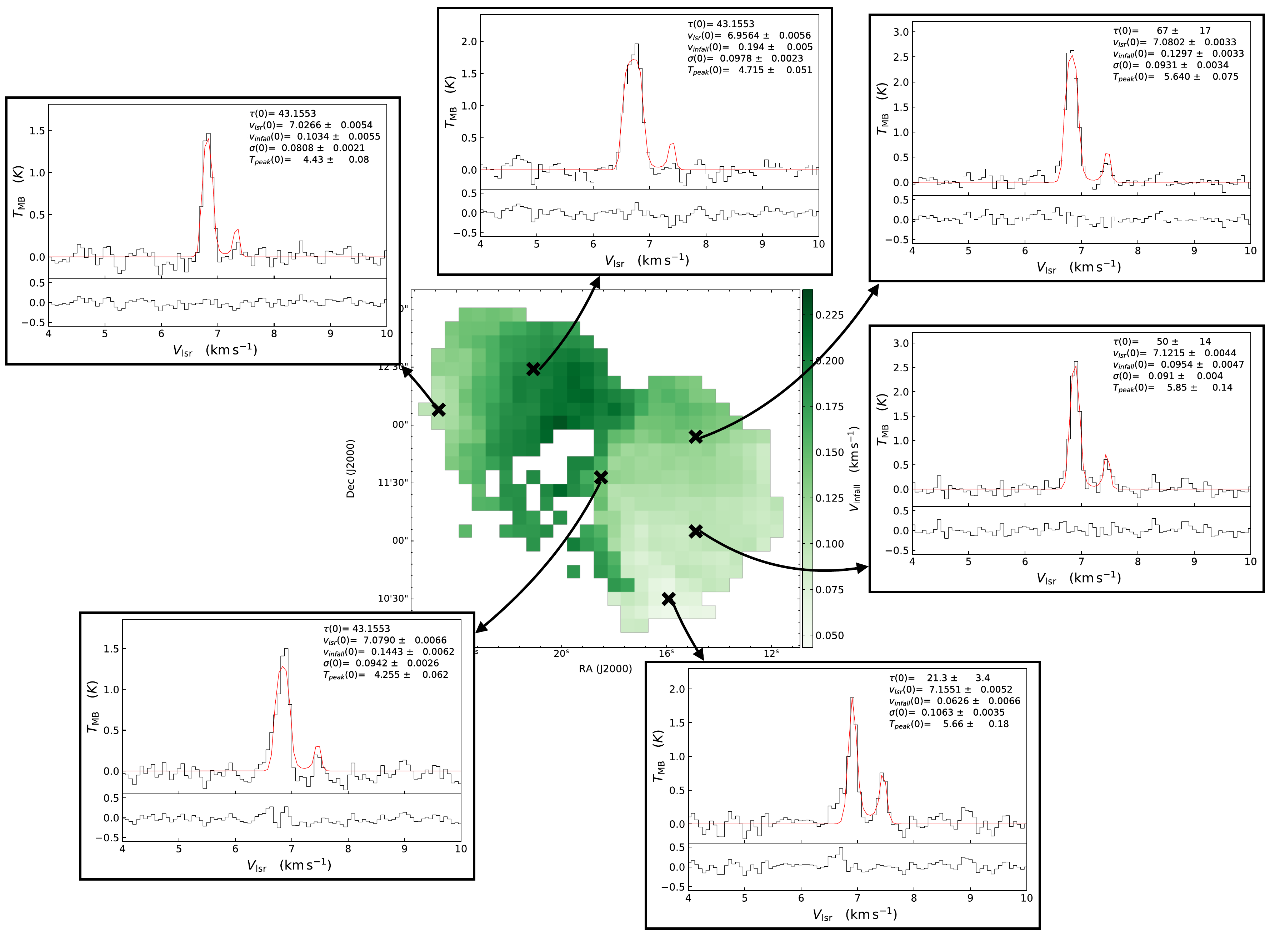}
      \caption{The central panel shows a zoom-in of the infall velocity map shown in Fig.~\ref{Hill5}. The small panels shows a selection of spectra at different position (black histograms), with overlaid the best-fit solution found with the Hill5 model (red curves). The best-fit parameters are reported in the top-right corners. The bottom sub-panels show the residuals of each fit. \label{Hill5_Spectra}}
   \end{figure*}

Figure \ref{Hill5_Spectra} shows a collection of a few \hcop (1-0) spectra and their associated best-fit found with the Hill5 model. At the bottom of each panel, the corresponding residuals are also shown. In general, the model performs well, and the residuals appear rather flat. An exception is the spectrum at the top-centre, which is taken from a region of the source where the red peak of the line is completely absorbed. In this case, the model is not able to reproduce this feature. In the bottom-right panel, instead, a blue wing can be seen. This has already been discussed at the end of Sect. \ref{spectra_descript}.




\end{document}